\documentclass[12pt]{article}

\usepackage{graphicx}

\title{ Field Correlator Method  for the   confinement   in QCD  }
\author{  Yu.A.Simonov \\
State Research
Center\\Institute of Theoretical and Experimental Physics, \\
Moscow, 117218 Russia}

\newcommand{\beq}{\begin{eqnarray}}
 \newcommand{\eeq}{\end{eqnarray}}
\newcommand{\be}{\begin{equation}}
 \newcommand{\ee}{\end{equation}}

 \def\la{\mathrel{\mathpalette\fun <}}
\def\ga{\mathrel{\mathpalette\fun >}}
\def\fun#1#2{\lower3.6pt\vbox{\baselineskip0pt\lineskip.9pt
\ialign{$\mathsurround=0pt#1\hfil ##\hfil$\crcr#2\crcr\sim\crcr}}}

\newcommand{{\SD}}{\rm SD}

\newcommand{{\Mc}}{\mathcal{M}}

\newcommand{\vex}{\mbox{\boldmath${\rm x}$}}

\newcommand{\ver}{\mbox{\boldmath${\rm r}$}}

\newcommand{\vep}{\mbox{\boldmath${\rm p}$}}

\newcommand{\veR}{\mbox{\boldmath${\rm R}$}}

\newcommand{\vek}{\mbox{\boldmath${\rm k}$}}
\newcommand{\ven}{\mbox{\boldmath${\rm n}$}}

\newcommand{\veB}{\mbox{\boldmath${\rm B}$}}

\newcommand{\veE}{\mbox{\boldmath${\rm E}$}}

\newcommand{\veal}{\mbox{\boldmath${\rm \alpha}$}}

\newcommand{\llan}{\langle\langle}
\newcommand{\rran}{\rangle\rangle}
\newcommand{\lan}{\langle}
\newcommand{\ran}{\rangle}

\begin{document}
\maketitle
\begin{abstract}
  The  theory of confinement based on the  stochastic field mechanism, known  as the Field Corrleator Method (FCM) is discussed in detail. Experimental and lattice data have accumulated a vast amount of material on the properties of confinement in QCD.
 We enumerate all these properties as 1)-7), and  discuss  beyond FCM two    existing approaches: 
  monopole  based   Dual Ginzburg-Landau (DGL) theory, and    Gribov-Zwanziger model, from this point of view.
 It is shown that the FCM  satisfies all required criteria. 
 We also prove its selfconsistency; in particular, it is shown that the string tension $\sigma$ is the only scaleful parameter in the theory beyond  fermion masses, and $\Lambda_{QCD}$ is  calculated explicitly to the lowest order in terms of $\sigma$.
  We also formulate physical consequences of   confinement, such as string breaking, Regge trajectories, role of confinement in the perturbation theory, chiral symmetry  breaking, confinement in the boosted systems etc. It is demonstrated that the FCM is  a suitable tool for the solution of these problems.
 \end{abstract}

 \section{ Introduction}
The problem of confinement and its internal structure remains an important issue nowadays, while this topic is studied in numerous papers for the last 45 years, starting  from the first papers \cite{1,2,3,4}. It was  generally assumed at first, that the most important role in confinement must play topologically nontrivial configurations, e.g. like magnetic monopoles \cite{2,3,4,5} or else other classical solutions: (multi) instantons, dyons, etc. The corresponding effective Lagrangians, establishing the form of dual Abrikosov fluxes \cite{6,7}, have soon been found \cite{8,9} and demonstrated possibility of the dual Ginzburg-Landau (DGL) theory. This topic is effectively elaborated till now, see \cite{10,11} for reviews.

Since the first definition of confinement via the area law of the Wilson loop \cite{1}, the lattice analysis of confinement plays the most important role,  which allows to define the most important properties of confinement and study this phenomenon quantitatively, see \cite{12}. These studies allowed to analyse the  QCD vacuum configurations and to search for monopole-like degrees of freedom, as it is done in the Abelian projection method (APM) \cite{13}, in the center vortex model \cite{14}, and the thick vortex model \cite{15}.

Meanwhile another approach  to confinement, based on the Field Correlator Method (FCM), was formulated in \cite{16}, see \cite{17,18,19} for reviews, which allowed to connect confinement directly to the (Gaussian) bilocal  field correlator $\lan F(x) F(y)\ran$, created in the QCD vacuum, and therefore sometimes called the Stochastic confinement (SC). One of the most crucial tests of this method is the analysis of  confinement between sources in different group representations  -- the so-called Casimir scaling (CS), which was done on the lattice \cite{20,21,22} and compared to the FCM predictions \cite{23,24}. In \cite{22} the agreement   was around 5\% for all  8 studied representations, which strongly  supported the FCM approach.

At the same time the analysis of CS in the DGL model \cite{8,25} has shown that the SU(3)  Casimir ratios cannot be reproduced for the fixed parameters of the model, i.e. for fixed values of the monopole mass $m_\chi$ and dual gauge field mass $m_B$, so that each representation  requires its own set of masses. Till now this discrepancy is not resolved and a reasonable modification of the DGL type or any other connected model, satisfying Casimir scaling for SU(3) or SU(N) groups, is not yet found, which  sets some limits on the presence of DGL configurations in the QCD vacuum.

This analysis can be prolonged  to take  into account the simple groups $F_4, E_6 $, and $G_2$, where linear confinement is present only up to some distance, see \cite{26,27,28,29,30,31,33}. Here the aim is to find the connection between the group structure and spatial and Casimir properties of confinement. As a basic point one can use here also the bilocal field correlators of the exceptional groups, which provide  Casimir scaling.

One of the most important issues of the CS is the proof of the  dominance of the bilocal correlator, which ensures CS, and the estimates of higher correlators, $\lan FFFF\ran$ etc. This analysis was first  done in \cite{34,35}, where it was shown that all properties of the correlators and, moreover, the quantitative expression for  bilocal correlator, i.e. the most part of the  confinement dynamics, can be  derived from the gluelump Green's functions, which were found analytically in \cite{36} and on the lattice in \cite{37}.

In this  way the theory of confinement has found its quantitative basis and can be called the FCM of confinement. In this theory the only parameter is  the string tension $\sigma$ (which may be expressed via $\Lambda_{QCD}$) and the field correlators are  expressed in terms of $\sigma$, which finally yields self-consistent connection  $\sigma (\sigma)$ \cite{34}. We consider this fact as the most important property of our approach, which is missed in the DGL type of models, where masses $m_B$ and $m_\chi$ and their ratio are not introduced till now  self-consistently.

All this and additional physical properties,  taken from lattice and hadron properties, can be formulated as necessary  properties of  the QCD  confinement mechanism, listed below.

\begin{enumerate}
\item Confinement is linear  in the $SU(N)$ field theory  for all measured distances, $R\la 1$ fm  as found on the lattice \cite{12,22}, whereas for QCD with $n_f>0$ at large distances linear confinement is flattening. Field correlators in QCD are exponentially damped at  large distances \cite{38,39,40, 41, 42}, $\lan F F\ran \sim \frac{c}{x^4} + d\exp (-\mu x)$.
\item Casimir scaling is found for all charge representations of $SU(3)$ with  O(5\%) accuracy in the range up to 1 fm  \cite{22}.
\item Flux tubes are observed  between the charges, with the radius, which is  slowly changing with distance between charges, see \cite{12,43}. A circular colormagnetic current is  observed  around flux tubes, asymptotically  satisfying the dual  London equation \cite{17,19}.  The excited flux tubes have the specific hybrid type spectrum \cite{20}. The $3q$ and $3g$ systems have the string  configuration of the  string junction and triangular type, respectively \cite{12}.
\item When going from static charges  to finite mass   fermions, one discovers the necessity of the scalar property of confinement, since otherwise the vector  confinement does not ensure  $q \bar q$ bound states \cite{44,45}.
\item Since we have the only scale in QCD, $\sigma$ or   $\Lambda_{QCD}$, which defines all quantities (in addition to quark masses),  the  confinement interaction should be expressed via $\Lambda_{QCD} \sim \sqrt{\sigma}$ as the only scale parameter.

\item The confinement theory should explain  the  interaction between  Wilson loops, observed on the lattice \cite{46}, in good agreement with FCM \cite{47}, and in particular, the visible lack of interaction between the Coulomb-like fluxes and colored flux tubes,  which ensures the nonscreening of the Coulomb interaction at large distances, observed on the lattice \cite{12,22} and in the hadron spectrum \cite{48}, and  explained in \cite{49} within the FCM.
 \item The confinement  theory should be applied to all known examples of field theory, e.g. it should yield no confinement for QED, linear confinement for all groups $SU(N)$, $N\geq 2$. It  can be also applied to other simple groups, like $G_2, F_4, H_6$, where linear confinement with $CS$ is present in the region $r< r_{\max}$ \cite{26,27,28,29,30,31,32,33}.
 \end{enumerate}

 Below we apply these properties as criteria to three types of confinement mechanisms, the FCM, the DGL theory, and the Gribov-Zwanziger \cite{50} approach. We also discuss the center vortex models \cite{14,15} and their possible connection to the FCM. 

 As we demonstrate below, using the concrete  gluelump structure of the field  correlators developed in \cite{34,35}, the FCM satisfies all criteria. This has allowed to calculate the confinement interaction at all distances, ensuring linear confinement for $r> \lambda \sim 0.1   $ fm, where $\lambda$ is  the inverse mass of the lowest gluelump, $M_{Glp}\approx 2$ GeV, calculated  via string tension $\sigma$.  This theory  was applied to the structure of flux tubes, originally in \cite{17,19} and recently   in \cite{49}.
 Surprisingly, our flux tubes confirm all the structure, observed on the lattice (\cite{12,22,43}), implying e.g. also the asymptotic  validity of dual London equation. This means, that the main  mechanism of dual  magnetic vacuum,  providing mass of propagating gluons, is present in our ``microscopic'' Gaussian--Gluelump  approach,  resembling in this respect   the macroscopic DGL approach.

It is remarkable that starting directly from  simplest (Gaussian) field correlators and not assuming any of DGL configurations, one arrives on  the microscopic  level (i.e. on distances $x\geq  \lambda \cong 0.2$ fm) at the field and current distributions specific for the macroscopic DGL equations.

Indeed, as we shall show below, for  the most general form of field correlators one obtains the circular color  magnetic  currents $\vek_D$ around flux tubes, which satisfy asymptotically the dual London's equation  $rot \vek_D= \lambda^{-2} \veE_D$ and  there emerges picture of  dual  superconducting fluxes around the dual Abrikosov string. In  all this picture no magnetic monopole d.o.f. are needed, the only microscopic reason of confinement is the  presence of the  scalar $D(z)$ in the vacuum correlator $\lan F_{\mu\nu} (x) \Phi F_{\lambda\sigma}(y) \ran \sim (\delta\delta-\delta\delta) D(x-y) +...$

The presence of such term in the QCD vacuum with or without quarks is proved by numerous lattice  calculations \cite{38,39,40,41,42}, and  in FCM it is calculated in a self-consistent way. Then one may ask oneself: why at all one should search for magnetic-monopole-like d.o.f. in the QCD configurations? Why one needs any topological configurations, since a simple scalar term $D(z)$ in the  microscopic correlator $\lan FF\ran$ already ensures the macroscopic dual  superconducting picture?

Or  in other words: what additional features of confinement are provided by the DGL type of theory?
As we shall see in the next chapters, one still has no explicit answer to this question in QCD, or SU(N) theories , where the FCM alone is sufficient to explain all known details of confinement till now,  but in more complicated theories like $G_2, F_4, H_6$, one may need other instruments, like the DGL or center vortex model. 

In what follows we shall shortly derive and discuss basic equations of our method (to be referred to as FCM ), demonstrate how it satisfies the conditions 1) -- 7), and find the points,  where other approaches fail.

The plan of  the  paper is as follows. In section 2 we give  the basics of the FCM and in section 3 discuss the properties  1)-7) within FCM. In section  4 the DGL approach is discussed   with respect to the   same  properties. In section 5 the approach of center vortex model (CVM) is shortly exposed and compared to the FCM approach.  In section 6 the necessary  scalar property of confinement is proved for light quarks. In section 7 we  discuss five features of the QCD dynamics, which are connected with  confinement, and demonstrate how the FCM is incorporated in resolving: the IR renormalon problem, chiral symmetry breaking, string breaking, confinement in  boosted systems, spin-dependent interaction etc. The concluding section gives the summary of results and discussion of possible development.

\section{Basics of the  Field Correlator Approach}

One starts with the expression of the Wilson loop \cite{1}, which can be also rewritten in terms of the field strength operators $F_{\mu\nu}$, using the nonabelian  Stokes theorem \cite{51} for the minimal surface $S_{\min} $ inside the   contour $C$
\be W(C) = \frac{1}{N_c} \lan tr P\exp(ig \int_C dz_\mu A_\mu (z) )\ran = \frac{1}{N_c} \lan tr P\exp(ig \int_{S_{\min}} d\sigma_{\mu\nu}  F_{\mu\nu} )\ran.\label{1}\ee

One can apply to (\ref{1}) the operator cluster expansion \cite{52}, which allows to expand in the exponent the connected terms, producing connected correlators $\llan\rran$.

Before  doing this one should define the gauge-covariant quantity, $\hat F (x) = \Phi (X,x) F(x) \Phi(x,X)$, where $\Phi(x,y) = P\exp ig \int^x_y A_\mu (z) dz_\mu$ is a parallel  transporter (pt), and we have chosen $x=X$ as an arbitrary common point to make all construction gauge invariant. Thus each $\hat F (x)$ is connected to $X$ by a pair of pt. Further on, in doing the vacuum averaging of products $\lan \hat F(1) \hat F(2)... \hat F(n) \ran$, one is exploiting the minimal action principle, which ensures that the main contribution to the average is given by the  configuration where all points 1,2,...$n$ are connected by pt's of minimal length, so that $\lan  \hat F(1) \hat F(2)= \lan\Phi(2,1)  F(1)\Phi(1,2) F(2)\ran$ and so on. As a result the vacuum averaging due to \cite{52} yields

$$ W(C) = \frac{1}{N_c} tr \exp \left[ - \frac{g^2}{2} \int d\sigma_{\mu\nu} d\sigma_{\lambda\rho}\llan F_{\mu\nu} F_{\lambda\rho}\rran + \right.$$
\be \left.+ \frac{g^4}{4!} \int d\sigma (1) d\sigma(2) d\sigma(3) d\sigma(4)  \llan \hat F(1)\hat F(2) \hat F(3)\hat F(4)\rran +O(g^6)\right].\label{2}\ee

We shall keep the first term  in the  exponent $O(g^2)$ as the basic approximation and later estimate other terms, following the discussion in \cite{18,23,24}. From \cite{51,52,19} one  can deduce that one can organize the connected clusters $\llan F(1) ... F(n)\rran$ in such a way, that all points $1,...n$ are connected by interaction, averaged in the vacuum averaging process. Assuming the   correlation length $\lambda$ for  this interaction, one obtains the estimate
\be I_n\equiv \int\int \llan F(1) ... F(n) \rran d\sigma (1) ... d\sigma (n) \approx f R T_4 (\lambda^2 f)^{n-1},\label{3}\ee
where $f$ is the order of magnitude estimate of the operator $F$.

The basic point of the FCM is  the expression for the vacuum field correlator, which in the colorelectric case is \cite{16}

$$  g^2 D^{(2)}_{i4k4} (x-y) \equiv \frac{g^2}{N_c} \lan tr_f (F_{i4} (x) \Phi(x,y) F_{k4} (y)
\Phi(y,x)\ran = (\delta_{ik} ) D^E(x-y)+$$ \be + \frac12 \left(
\frac{\partial}{\partial x_i} [h_k + {\rm ~perm}]\right) D_1^E (x-y), ~~
h_\lambda = x_\lambda -y_\lambda, ~~(x-y)^2 = \sum^4_{\lambda=1} (x_\lambda-y_\lambda)^2.
\label{4}\ee

Insertion of (\ref{4}) into (\ref{2}) yields the area law of the Wilson  loop

\be W(C) =\exp (-\sigma RT_4), ~~\sigma =\frac12 \int d^2z D^E(z).\label{5}\ee
Comparing (\ref{4}) with (\ref{3}), one can see that the estimate  holds
\be
\sigma \approx f^2\lambda^2, ~~  I_n \sim \frac{RT_4}{\lambda^2} (\sigma \lambda^2)^{n/2}, ~~ \frac{I_4}{I_2} \sim \sigma \lambda^2\label{6}\ee

 In the Appendix 1 we estimate in detail the quartic correlator, supporting validity of Eq. (\ref{6}).

We now turn to the calculation of static potentials generated by $D^E,   D_1^E$.

We start with the Wilson loop, $W(C) = \exp  \left(-\frac{g^2}{2} \int d\sigma \int d\sigma \lan F  F\ran\right)= \exp  \left (-\int (V_D^R) + V_1(R))dt_4\right)$  and consider an interval $\Delta t_4>\lambda$ in both integrals $ \int dt_4$ and $\int d \sigma (u) \int  d\sigma (v)$, which yields  (see Fig.1), $d\sigma (u) = du_4 du_1$.


\begin{center}

\begin{figure}
\begin{center}
  \includegraphics[width=4cm, ]
  {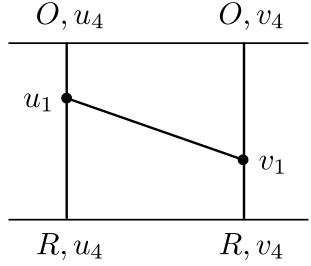}

\end{center}
  \caption{ Calculation of the static potentials $V_D, V_1$ from field correlators}

\end{figure}
 \end{center}

$$ V_D (R) \Delta t_4 =  2\int^R_0 du_1 \int^{\Delta  t_4}_0 du_4 \int^R_0 dv_1 \int^{\Delta  t_4}_0 dv_4 D^E(u-v)=$$
$$ =2 \int  d \frac{u_1+v_1}{2} \int  d(u_1-v_1) \int d \frac{u_4+v_4}{2} \int d(u_4-v_4)  D^E (u-v)=$$
\be = \Delta t_4 2 \int^R_0 (R- w_1) dw_1 \int^{\Delta t_4}_0 d w_4  D^E \left(\sqrt{w^2_1 + w^2_4}\right).\label{8g}\ee

As a result for $\Delta t_4\gg \lambda$, one obtains the static potential  $V_D (R)$  for the fundamental  charges
\be V_D (R) = 2 \int^R_0 (R-w_1) dw_1 \int^\infty_0 dw_4 D^E  \left(\sqrt{w^2_1+ w^2_4}\right)= V_{\rm conf} (R) + V_D^{\rm sat} (R), \label{9g}\ee
where $V_D^{\rm sat}(R)$ is negative and saturates at large $R$.

For  the charge  representation $D$ the Gaussian correlator $D^E$ defines the interaction between static charges in the representation $D=3,8,6,...$\be
V_D (R) = C_D \int^R_0 (R-w_1) dw_1
\int^\infty_0 dw_4 D^E  \left(\sqrt{w^2_1+ w^2_4}\right)     , \label{7}\ee
where $C_D = 2\frac{C_2(D)}{C_2(f)}$, and $C_2 (D)$  is the quadratic Casimir coefficient for the representation $D$ \cite{16,18}.

In a similar way one obtains static potential $V_1(R)$, generated by $D_1^E (z)$, \cite{17,18}
 \be V_1 (R) = \int^R_0 w_1 dw_1 \int^\infty_0 dw_4 D_1^E \left(\sqrt{w^2_1 + w^2_4}\right)\label{10g}\ee

As it is shown in Appendix 2, after regularization   one arrives at the final form of $V_1(R)$ in (\ref{10g})
\be V_1(R) = - C_2 \frac{\alpha_s}{R} + V_1^{\rm sat}(R) .\label{13g}\ee
Since all  components  of $V_D, V_1$ are proportionals to the   quadratic field correlator, the potential for any representation $D=f,$ adj, ... are proportional to the coefficient $C_2 (D)$, where $C_2(D)$ is  the Casimir factor.

The most important step was done in \cite{34,35}, where $D^E_1$ and  $D^E$ were expressed via the one- and  two-gluon gluelump   Green's function, as shown in Figs. 2 and 3, respectively.

\begin{center}

\begin{figure}
\begin{center}
  \includegraphics[width=6cm, ]{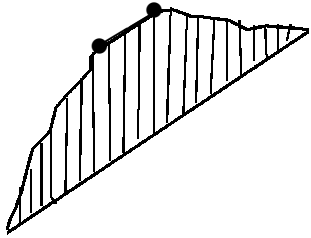}

\end{center}
  \caption{The one-gluon  gluelump Green's function }

\end{figure}
 \end{center}


\begin{figure}
\begin{center}
\includegraphics[width=60mm,keepaspectratio=true]{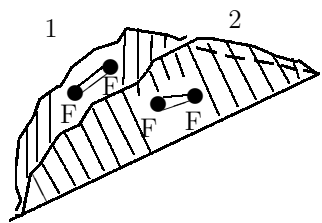}
\end{center}
\caption{The two-gluon  gluelump Green's function }
\end{figure}

 Indeed, writing $F_{\mu\nu} $ in (\ref{4}) as $F_{\mu\nu} = \partial_\mu A_\nu -\partial_\nu A_\mu + ig [A_\mu, A_\nu],$ one obtains for $g^2 D^{(2)}_{\mu\nu, \lambda\sigma}  $ in (\ref{4}) the derivative terms, denoted as $D_{1, \mu\nu\lambda \sigma}$  and $O(g^4)$ terms proportional to $\lan [A_\mu, A_\nu] \Phi (x,y)  [A_\lambda, A_\sigma]\ran \equiv  G^{(2g)}$.
 It is clear, that $D_{1\mu\nu\lambda\sigma}$ contain the term $\lan A_\mu (x) \Phi (x,y) A_\lambda(y)\ran = G^{(1g)} (x,y)$ which is the one-gluon-gluelump Green's function, while $G^{(2g)}$ is the two-gluon gluelump Green's function, calculated in \cite{36,37}.
 A more detailed derivation  is discussed in Appendix 2. As a result  one can associate $D_1^E$  with the derivative terms  in  correlator $  \lan F \Phi  F \ran$, namely,

\be D_{1\mu\nu, \lambda\sigma} (x,y) = \frac{g^2}{2N^2_c} \left\{\frac{\partial}{\partial x_\mu} \frac{\partial}{\partial y_\lambda} \lan tr_a A_\nu (x)  \Phi(x,y) A_\sigma (y) \ran  + {\rm perm}\right\},\label{11g}\ee
and    $D_1$ is expressed via the    one-gluon  gluelump Green's function with the asymptotics found in \cite{34,35}
\be D_1^E (x) =- \frac{2g^2}{N^2_c} \frac{d G^{(1g)} (x)}{ dx^2},  D^E_1 (x) \approx \frac{A_1}{|x|} e^{-M_1|x|},\label{12g}\ee
where $A_1 = 2C_2 \alpha_s   \sigma_{\rm adj} M_1$.

One can see that $gg\Phi$ is  the bound state of two gluons with a static gluon from $\Phi(x,y)$, so that in the transverse plane the $gg\Phi$ configuration looks like   a triangle , where straight lines of its sides  represent confining strings between adjoined charges. A similar consideration for the $O(g^4)$ terms yields $D^E(x-y)$.  \be D^E(x-y) =
\frac{g^4 (N_c^2-1)}{2} G^{(2g)} (x,y)= g^4 N_c C_2 (f)  G^{(2g)}(x,y).\label{8}\ee

The spectrum of $G^{(2g)}$ was found both analytically \cite{36} and on the lattice \cite{37}, and in \cite{34,35} the asymptotics was found as

\be G^{(2g)} (x)  \approx 0.108 ~\sigma^2_f
e^{-M_0^{(2g )}|x|},~~ x\ga (M_0^{ (2g )} )^{-1} \label{9}\ee
while $M_0^{(2g)} \approx 2$ GeV,  $ (M_0^{ (2g )} )^{-1} \equiv \lambda_0 =0.1  $ fm. The  mixing of $M^{(2g)}$ and $M^{(1g)} $ and the account of color Coulomb interaction  imply the lowering of $M^{(2g)}$, and  hence the increasing of $\lambda, \lambda\la 0.2$ fm.  Note here, that $M_0^{(2g)}$ is expressed via $\sigma$, namely, according to \cite{36,37}.
\be M_0^{(2g)} \cong (4\div 4.5) \sqrt{\sigma_f}\label{10}\ee
and it is  the large ratio $4\gg 1 $, that ensures the small ratio of  $\frac{I_n}{I_2}$ in (\ref{6}), e.g.
\be\frac{I_4}{I_2} \sim \frac{\sigma}{(4.5 \sqrt{\sigma})^2} \sim \frac{1}{20}.\label{11}\ee

This result will be basic for the properties 1) and 2), listed in the Introduction.

One can compare this result for $\lambda$ with direct lattice measurements of field correlators \cite{38,39,40,41,42}, which yield $\lambda\approx 0.2$ fm.

We are now coming to the most  important property of the FCM mechanism of confinement -- the scale selfconsistency. Indeed, in FCM (and in QCD with massless quarks  in general) one has the only nonperturbative scale, which defines 99\% of mass in the visible part of the the universe. It can be chosen as a scale since confinement and $\sigma$ explain the  nucleon masses. Now from the definition of $\sigma$ (to  the lowest order in $\alpha_s)$ $\sigma_f = \frac12 \int D^E (x) d^2 x$,  and the asymptotic expressions (\ref{8}), (\ref{9}) for $D^E(x)$ one obtains the selfconsistency condition
\be \sigma_f \la \pi\lambda^2 \cdot 0.0108\cdot 8 \pi^2 \cdot \alpha^2_s (N^2 _c-1) \sigma^2_f.\label{18a}\ee (The sign $<$ is due to overestimating  $D^E(x)$ using asymptotics (\ref{9}), since $D^E$ is smaller for $x\to 0$, and  has a minimum at $x=0$, as shown in Appendix 3). Here $\lambda =1/M$ and we can associate the momentum scale of $\alpha_s$ with  the  gluelump mass $M\cong 2$ GeV $\cong 5 \sqrt{\sigma}$, and from (\ref{18a}) one finds the dependence $\alpha_s (M)$
\be \alpha^2_s (M)\equiv(\alpha^*_s)^2 = \frac{M^2\cdot 0.037}{\sigma_f (N^2_c-1)}\cong 0.104, ~~ (N_c=3,~~ \alpha_s^* =0.322,~~ M=2~{\rm GeV}) \label{18a'}\ee
on the other hand one can use the one-loop approximation for $\alpha_s (M)$ with the IR correction found earlier (see Appendix 3 for details).

\be \alpha_s (M) = \frac{4\pi}{\beta_0 \ln \left( \frac{M^2+ M^2_B}{\Lambda^2}\right)}, ~~ M_B \cong 2\pi\sigma_f \approx 1~{\rm GeV}\label{18''}\ee
which yields (taking into account the sign $<$ in (\ref{18a})) for $N_c=3$

\be \Lambda_\sigma \ga\sqrt{M^2 +M^2_B}\exp \left(- \frac{4\pi}{\beta_0\alpha_s^*}\right) \cong 270~{\rm MeV}\label{18a'''}\ee

As one can see in (\ref{18a'''}) the $N_c$ dependence in $\alpha^*_s\beta_0$ is  compensated at large $N_c$, and the limiting value of $\Lambda_\sigma(N_c=\infty)$ is equal to 340 MeV, and as discussed in Appendix 3, Eq. ({A3.5}) should be divided by 1.3 to compare favourably  with $\Lambda_{QCD}^{\overline{MS}} $. The resulting values of $\alpha_s (2 $ GeV) and $\Lambda_{QCD}^{\overline{MS}}$ are well within the PDG limits.

In  this way we have  expressed to the lowest order the $\Lambda_{QCD}$ via the string tension, which can be now considered as the only scale constant in QCD beyond the fermion masses.

We turn now to  the phenomenon of flux tubes and their internal structure. It is a widespread notion that flux tubes are necessary and unique result of the DGL theory, producing a dual magnetic flux in the medium filled by the Higgs -- like monopole condensate. However one obtains the similar picture of a flux tube  directly from the quadratic field correlators and without any additional parameters except $\sigma$ and $ \lambda=\frac{1}{c\sqrt{\sigma}}, ~ c\approx 4$.

Indeed, following \cite{19,49}, one can measure the field $F_{\mu\nu}$ produced in the contour $C$, as shown in Fig. 4 (the so-called connected probe) and write


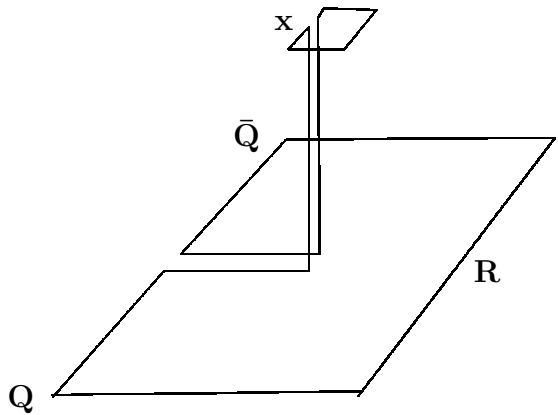
\begin{figure}[h] 
\setlength{\unitlength}{1.0cm}
\centering
\unitlength 1mm 
\linethickness{0.5pt}
\ifx\plotpoint\undefined\newsavebox{\plotpoint}\fi
\begin{picture}(105,90)(0,0)
\multiput(20.25,38.25)(.033675799,.038242009){438}{\line(0,1){.038242009}}
\put(35,55){\line(1,0){19.25}} \put(54.25,55){\line(0,1){32.5}}
\multiput(54.25,87.5)(-.0335366,-.0365854){82}{\line(0,-1){.0365854}}
\put(51.5,84.5){\line(1,0){7.5}}
\multiput(59,84.5)(.03373016,.04166667){126}{\line(0,1){.04166667}}
\multiput(63.25,89.75)(-.875,.03125){8}{\line(-1,0){.875}}
\multiput(56.25,90)(-.032609,-.054348){23}{\line(0,-1){.054348}}
\multiput(55.5,88.75)(.03125,-3.9375){8}{\line(0,-1){3.9375}}
\put(55.75,57.25){\line(-1,0){18.5}}
\multiput(37.25,57.25)(.03373494,.036746988){415}{\line(0,1){.036746988}}
\multiput(51.25,72.5)(4.46875,.03125){8}{\line(1,0){4.46875}}
\multiput(87,72.75)(-.0336970475,-.0442875481){779}{\line(0,-1){.0442875481}}
\multiput(61.25,39)(-2.75,-.033333){15}{\line(-1,0){2.75}}
\put(51,88.0){\makebox(0,0)[cc]{$\mathbf{x}$}} \put(78,55){\makebox(0,0)[cc]{$\mathbf{R}$}}
\put(16,38){\makebox(0,0)[cc]{$\mathbf{Q}$}}
\put(46,73){\makebox(0,0)[cc]{$\mathbf{\bar Q}$}}
\end{picture}
\vspace{-3.5cm}
\caption{The connected probe  for measuring  color field in the flux tube}
\vspace{-0.1cm}
\label{fig:fig01}
\end{figure}
\vspace{0.5cm}

 \be
F_{\mu\nu} (x) = \int_S d\sigma_{\alpha\beta} (y) g^2 D^{(2)}_{\alpha\beta\mu\nu} (x-y), \label{13}\ee
where $D^{(2)}$ is given in (\ref{4}). Writing $D^E(z)$ for simplicity as
\be D^E(z) = \frac{\sigma}{\pi\lambda^2} \exp \left(-\frac{|z|}{\lambda}\right), \label{14}\ee (which satisfies both (\ref{9}) with $\alpha_s \approx 0.2$ and (\ref{5})), the  mixing of $M^{(2g)}$ and $M^{(1g)} $ and the account of color Coulomb interaction  imply the lowering of $M^{(2g)}$ and increasing of $\lambda = \frac{1}{M^{(2g)}}$. In  all applications to  flux tubes it is convenient to choose $\lambda$ around the value of $\lambda=0.2 $ fm.  From (\ref{13}) one obtains the colorelectric field in the flux tube \cite{19,49}
\be \veE^D  = \ven \frac{2\sigma}{\pi} \int^{R/\lambda}_0 du \left| u\ven- \frac{\ver}{\lambda}\right| K_1 \left(\left| u\ven- \frac{\ver}{\lambda}\right|\right)\label{15}\ee
where $\ven$ is along the flux tube. In a similar way one defines the magnetic current $\vek^D = rot \veE^D$, and at the midpoint   between charges and  at distance $r_\bot$ from the axis it is equal  to
\be
\vek^2_D (r_\bot) = \frac{4\sigma^2 r^2}{\pi^2\lambda^4} \left( \int^{\frac{R}{2\lambda}}_{-\frac{R}{2\lambda}} dx K_0 \left( \sqrt{x^2+\frac{r^2_\bot}{\lambda^2}}\right)\right)\label{16}\ee

As was shown in \cite{19}  the dual London equation:  $rot \vek = \lambda^{-2} \veE$ is satisfied by (\ref{15}), (\ref{16}) asymptotically at  $r_\bot \to \infty$, see  Fig. 5.

\begin{figure}[h]
\setlength{\unitlength}{1.0cm}
\begin{center}
\begin{picture}(9.2,8.5)
\put(0.3,0.3){\includegraphics[height= 6cm]{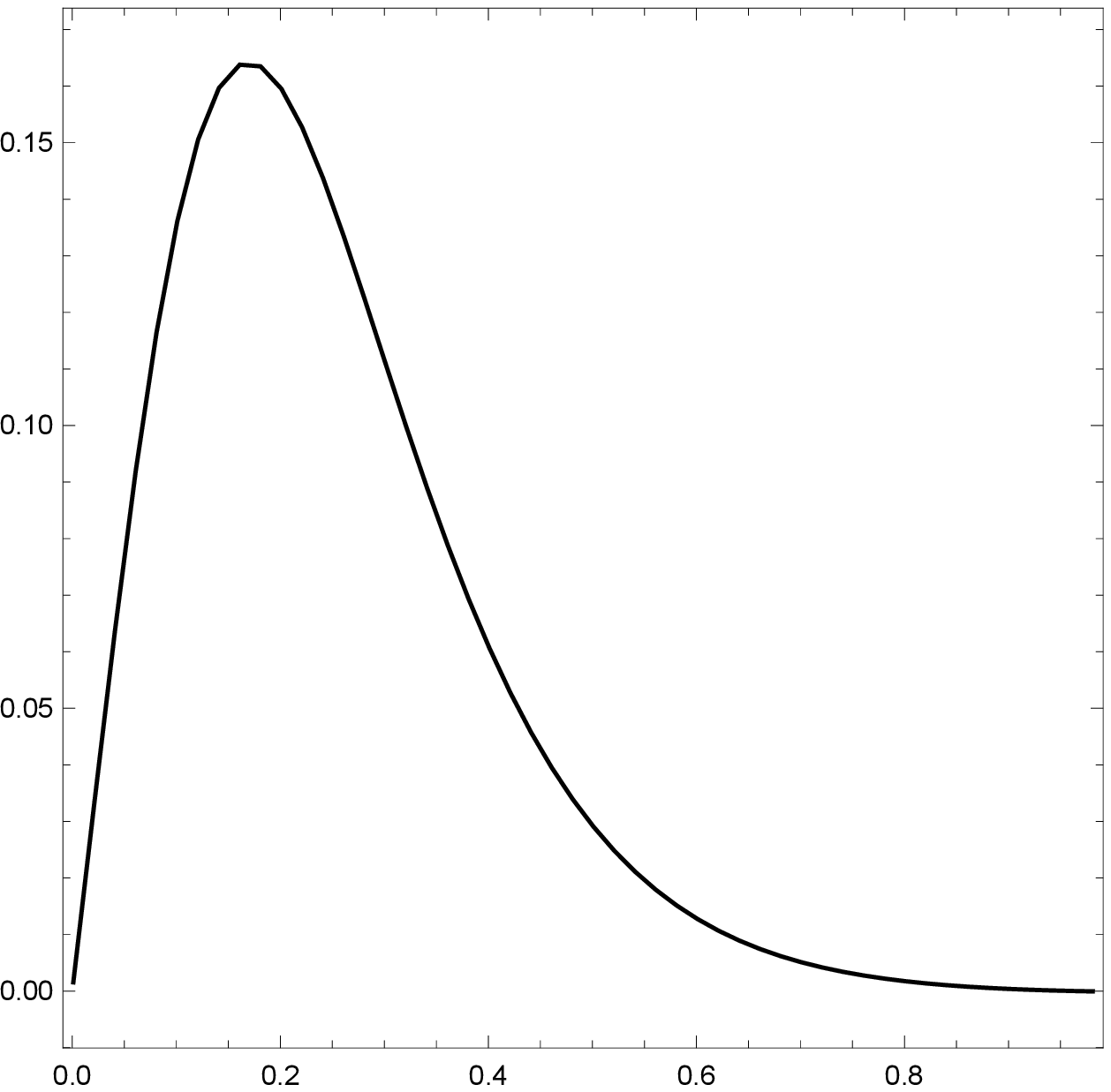}}
\put(0.5,7.95){\makebox(0,0)[cc]{{$\mathbf{k}$}}}
\put(8.2,0.2){\makebox(0,0)[cc]{$r_{\perp}$}}

\put(0.0,4.0){\rotatebox{90}{GeV$^3$}}
\put(4.6,0.2){\makebox(0,0)[cc]{fm}}
\end{picture}
\caption{The transverse radius dependence of the CM current }
\label{fig:fig10}
\end{center}
\end{figure}

To complete this picture one also calculates colorelectric fields due to correlator $D_1^E$, as it is done in the Appendix 2, with the resulting  field $\veE_1 $, depending on the same parameter $\lambda$. As a result one obtains the profiles of the flux tube for different distances $R$ between charges, shown in Fig.6, 7, 8,9.

One can  see in Figs. 6,7,8,9 a rather stable profile, only weakly  depending  on $R$, in good agreement with lattice data \cite{43}.


\begin{figure}[h]
\setlength{\unitlength}{1.0cm}
\begin{center}
\begin{minipage}[h]{0.48\linewidth}
\begin{picture}(7.25,7.0)
\put(0.3,0.3){\includegraphics[height=5cm]{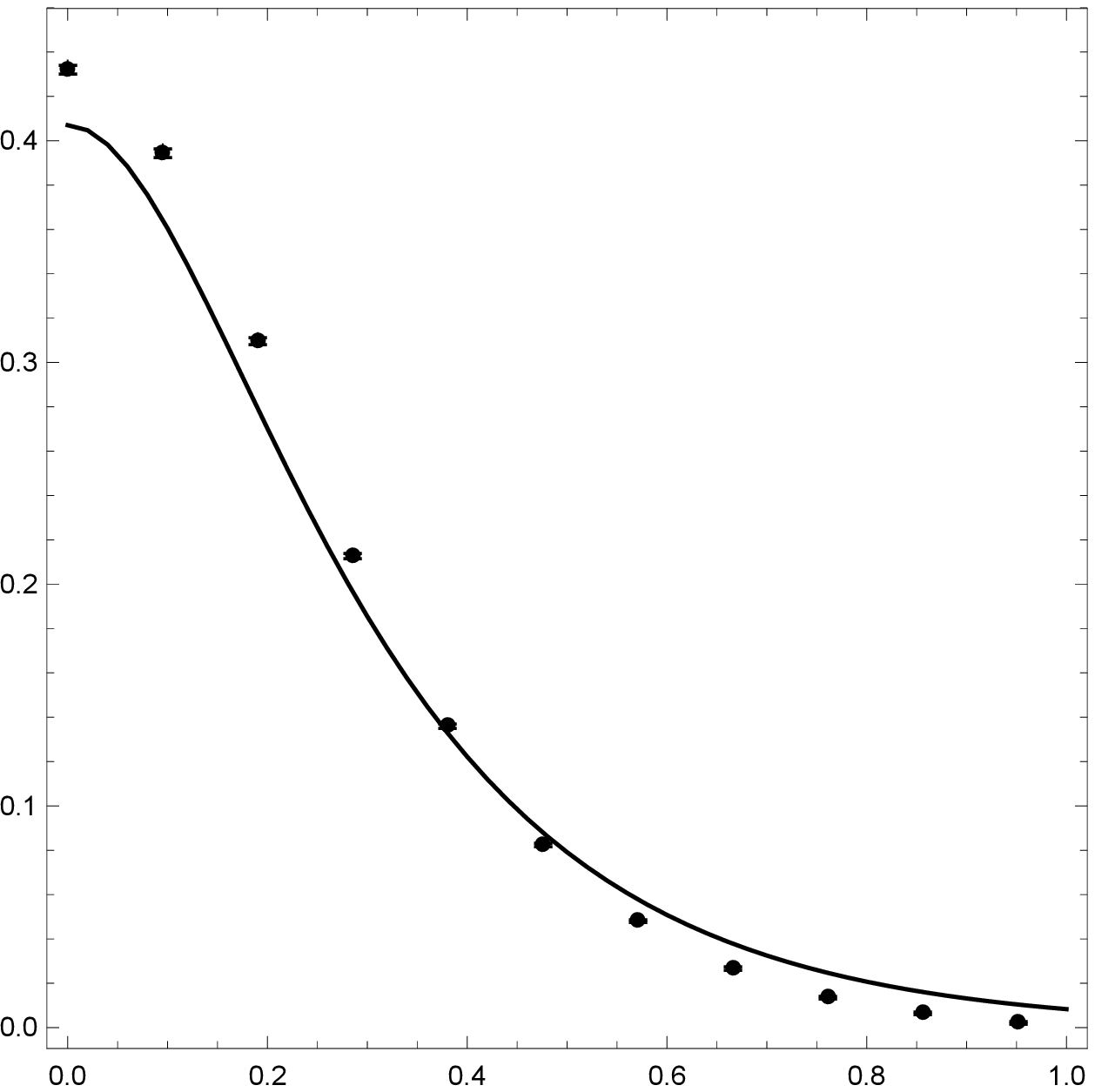}}
\put(0.25,7.0){\makebox(0,0)[cc]{{$E_3$}}}
\put(7.2,0.0){\makebox(0,0)[cc]{$r_{\perp}$}}

\put(-0.1,3.3){\rotatebox{90}{GeV$^2$}}
\put(3.95,0.15){\makebox(0,0)[cc]{fm}}
\end{picture}
\caption{$E_{3}=E_{3}(r_{\perp},\,R=0.76{\rm fm})$. The transverse radius dependence of the CE field strength for the fixed flux tube length $R=0.76 {\rm fm}$. The dots with error bars are from the lattice measurements in \cite{43}.}
\label{fig:fig06}
\end{minipage}
\hfill
\begin{minipage}[h]{0.48\linewidth}
\begin{picture}(7.25,7.0)
\put(0.3,0.3){\includegraphics[height=5cm]{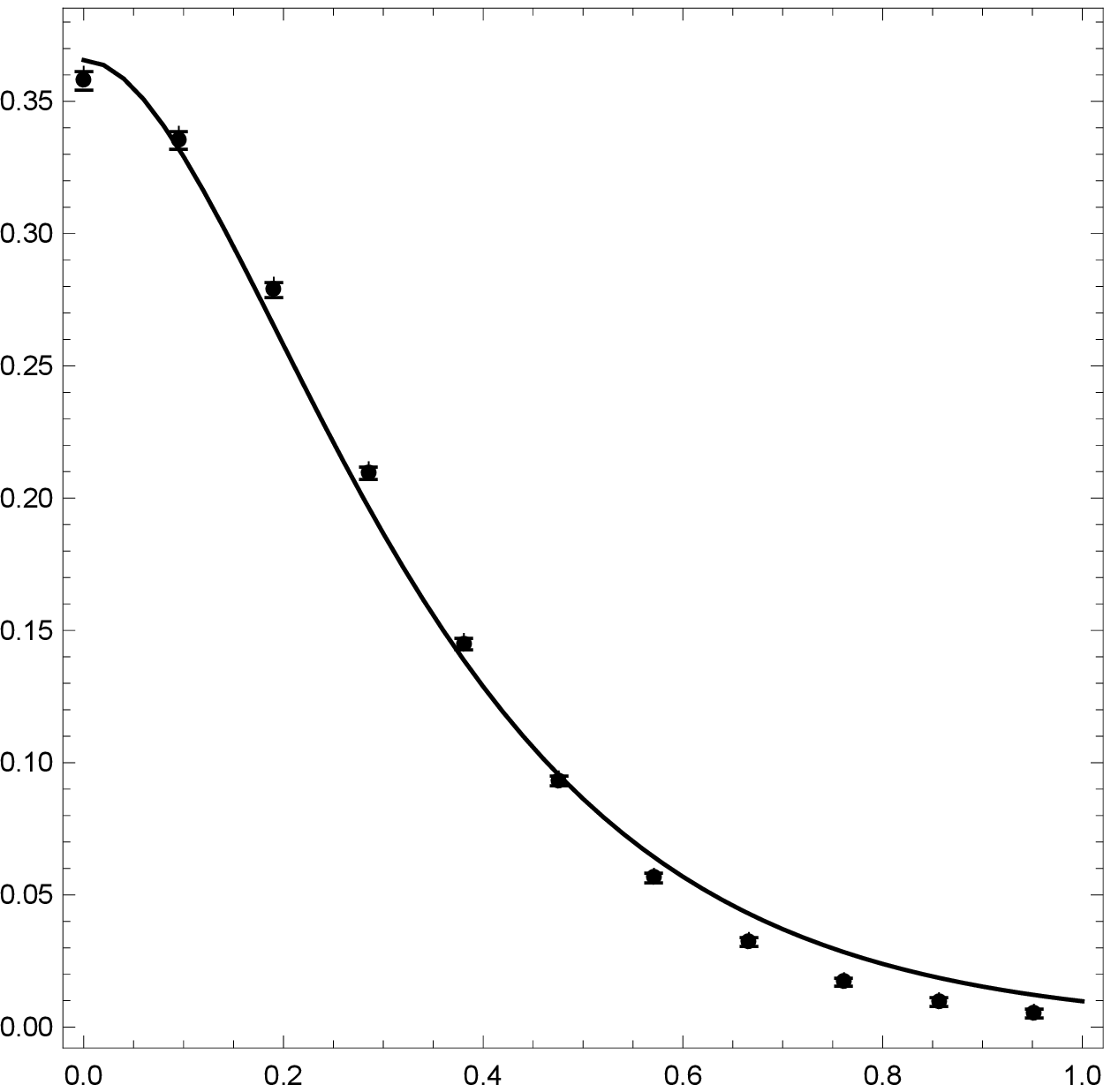}}
\put(0.25,7.0){\makebox(0,0)[cc]{{$E_3$}}}
\put(7.2,0.0){\makebox(0,0)[cc]{$r_{\perp}$}}

\put(-0.1,3.3){\rotatebox{90}{GeV$^2$}}
\put(3.95,0.15){\makebox(0,0)[cc]{fm}}
\end{picture}
\caption{$E_{3}=E_{3}(r_{\perp},\,R=0.95 {\rm fm})$. The transverse radius dependence of the CE field strength for the fixed flux tube length $R=0.95 {\rm fm}$. The dots with error bars are from the lattice measurements in \cite{43}.}
\label{fig:fig07}
\end{minipage}
\end{center}
 \end{figure}

 \begin{figure}[h]
\setlength{\unitlength}{1.0cm}
\begin{center}
\begin{minipage}[h]{0.485\linewidth}
\begin{picture}(7.25,7.0)
\put(0.3,0.3){\includegraphics[height= 5cm]{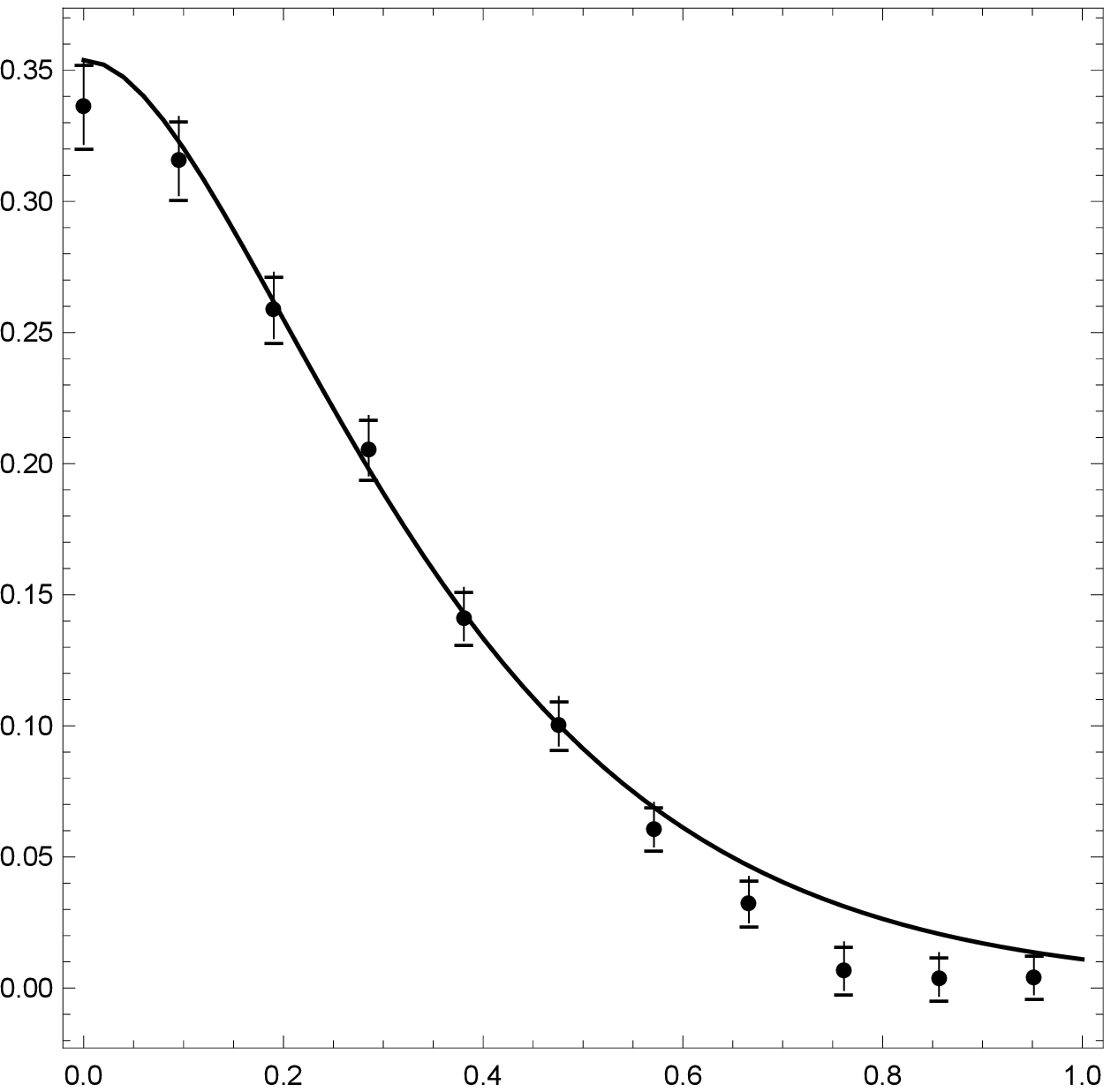}}
\put(0.3,7.2){\makebox(0,0)[cc]{{$E_3$}}}
\put(7.2,0.0){\makebox(0,0)[cc]{$r_{\perp}$}}

\put(-0.1,3.3){\rotatebox{90}{GeV$^2$}}
\put(3.95,0.15){\makebox(0,0)[cc]{fm}}
\end{picture}
\caption{$E_{3}=E_{3}(r_{\perp},\,R=1.14 {\rm fm})$. The transverse radius dependence of the CE field strength for the fixed flux tube length $R=1.14 {\rm fm}$. The dots with error bars are from the lattice measurements in \cite{43}.}
\label{fig:fig08}
\end{minipage}
\hfill
\begin{minipage}[h]{0.485\linewidth}
\begin{picture}(7.25,7.0)
\put(0.3,0.3){\includegraphics[height= 5cm]{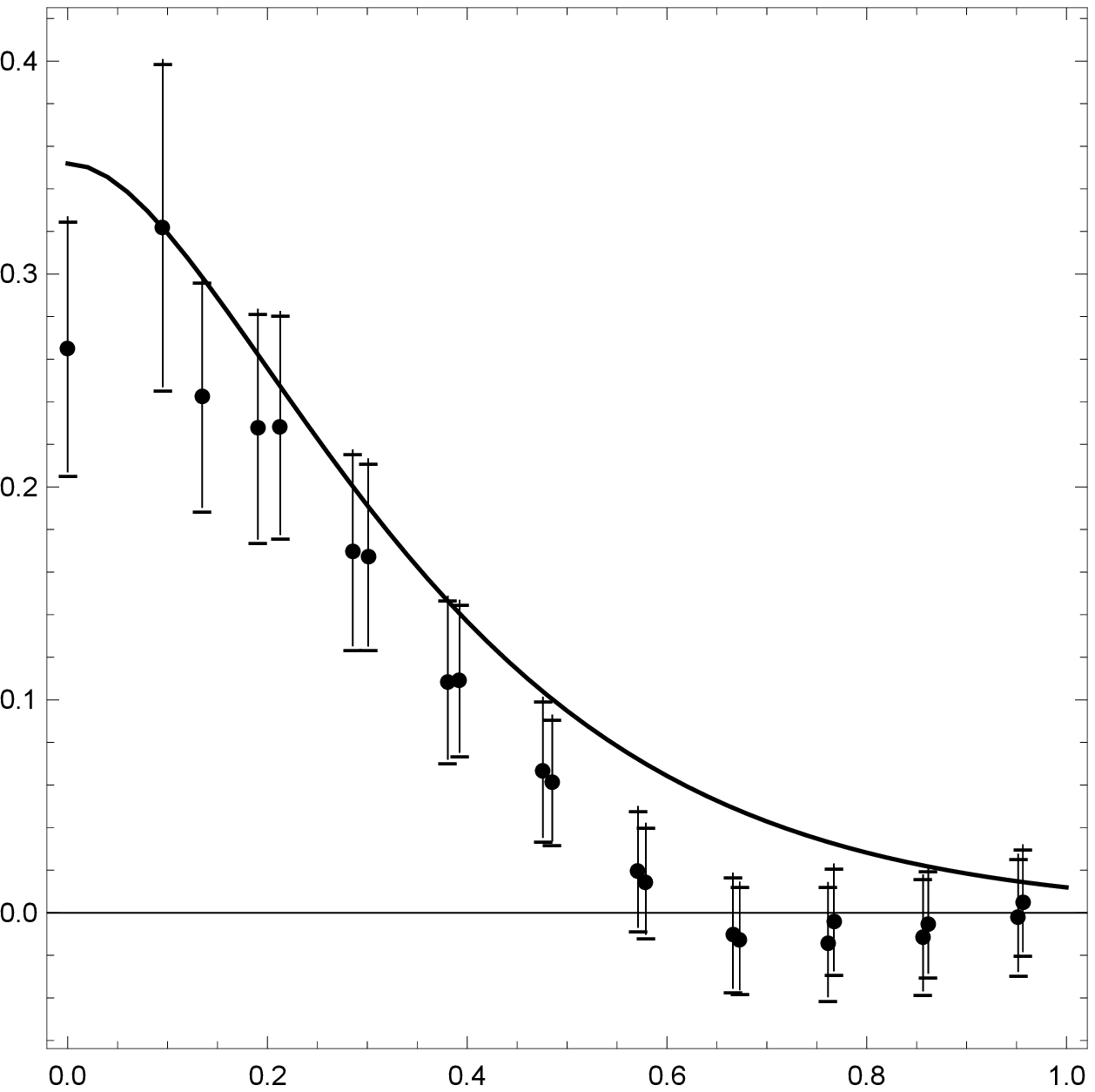}}
\put(0.3,7.2){\makebox(0,0)[cc]{{$E_3$}}}
\put(7.2,0.0){\makebox(0,0)[cc]{$r_{\perp}$}}

\put(-0.1,3.3){\rotatebox{90}{GeV$^2$}}
\put(3.95,0.15){\makebox(0,0)[cc]{fm}}
\end{picture}
\caption{$E_{3}=E_{3}(r_{\perp},\,R=1.33 {\rm fm})$. The transverse raius dependence of the CE field strength for the fixed flux tube length $R=1.33 { \rm fm}$. The dots with error bars are from the lattice measurements in \cite{43}.}
\label{fig:fig09}
\end{minipage}
\end{center}
\end{figure}


As was shown above, at the basis of FCM is the property of the Wilson loop, and hence one can immediately derive the effects of confinement, writing any amplitude in terms of Wilson loops. Above  we have considered only the simplest case, when all interaction inside $W(C)$ is of $np$ character. In the Appendix 2 we have shown that the correlator $D^E_1$ generates the instantaneous  Coulomb interaction $V_c(r)$ plus vector-like interaction $V_1^E$,  entering the Polyakov loop, $L_f =\exp \left( - \frac{V_1^{(E)}(\infty)}{2 T}\right)$.

In this way the color  Coulomb and confinement  interaction enter additively the total instantaneous potential, as it is supported by lattice data \cite{12,21,22}.

However, one  should consired these ``dynamical'' valence gluons in the confining film. In this case the main point is, how it interacts with the confining film and whether it produces the screening effect in the gluon-exchange interaction. This point was studied in \cite{48}, where it was shown  that the  resulting screening is small, with $\mu_{scr}\la 0.2$ GeV for light quarks and screeninig is not seen in heavy quarkonia up to distances $\sim 1.2$~fm. The  situation, when gluon  exchange is considered within the confining film, is shown in Fig.10, and in  \cite{49} it was explained why the screening is strongly damped.

\begin{figure}[h] 
\setlength{\unitlength}{1.0cm}
\centering
\unitlength 1cm 
\linethickness{0.4pt}
\begin{picture}(10.0,6.0)(0,0)
\put(0.1,0.1){\includegraphics[height=5.0cm]
{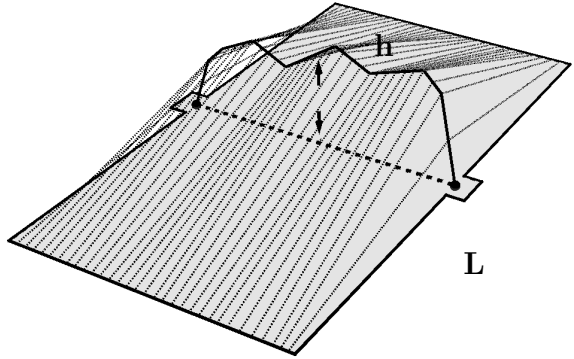}}
\put(5.3,4.4){\makebox(0,0)[cc]{$\mathbf{h}$}} \put(6.5,1.5){\makebox(0,0)[cc]{$\mathbf{L}$}}
\end{picture}
\vspace{-0.1cm}
\caption{The minimal area surface for the gluon exchange interaction.}
\vspace{-0.1cm}
\label{fig:fig03}
\end{figure}\vspace{1.0cm}

The  principle  of the minimal area for the surface of the Wilson loop operates also for the more complicated objects, like 3q baryons and 3g glueballs, which accordingly have the Wilson loops based on string junction and triangle forms, shown in Fig.11, see \cite{19} for details and the full theory of 3q baryons in \cite{53*}.

\begin{figure}[!t]
\hspace*{-0.5cm}
\centering
{\includegraphics[height=5.0cm]
{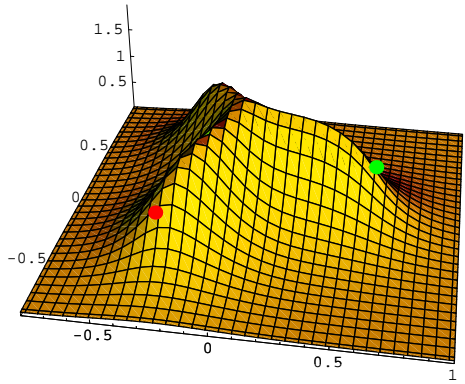}}
\caption{A distribution of the field $\vec E^{(B)}$
in GeV/fm with the only correlator  $D$ contribution considered in the
quark plane for equilateral triangle with the side 1 fm.
Coordinates are given in fm, positions of quarks are marked
by points. }
\end{figure}

Finally we touch on important point of excited QCD strings, which can be treated  in the lattice measurements of hybrid states \cite{53}, lattice measurements of flux tube excitations \cite{54}, and finally in the FCM theory \cite{55}. In the last case the FCM theory predicts excitations of the QCD string in the form of  an additional gluon, ``sitting on the string'' i.e. in the form of local excitation leading to the vibration of the string, with an  explicit probability amplitude for the space-time position of the gluon, while in the standard  string theory one  considers the string as a whole. From this point of view the analysis and comparison of the results of \cite{ 53,54,55} is of vital importance.

\section{The FCM theory $vs$ properties 1)--7)}

1. We start with the property 1), stating the observation of the linear  confinement in the whole measured region, 0.1 fm $<R<1$ fm,  \cite{21,22,12}. As it is clear from (\ref{7}), (\ref{8}), (\ref{9}), taking into account that $D^E (u,v) = D^E \left(z = \sqrt{u^2+v^2}\right)$ is essentially nonzero only in the region $r\la\lambda\approx 0.2$ fm, that the potential $V_D(r)$ has a linear behavior for all $r\ga \lambda$ in agreement with all lattice data \cite{12,21,22} and the  results in quarkonium  structure \cite{57}. This refers to all groups  $SU(N_c) , N_c \geq 2$, however, one should have in mind that in the higher $O(g^n), n\geq 4$ orders  or in the nonperturbative ($np$) string breaking mechanism   the  adjoint string may break. This phenomenon of string breaking into two gluelumps  brings in the flattening of the confining potential for adjoint and higher $D$ charges, which starts  at larger values of $r$  $r\ga r_{\max}$ because of larger gluelump mass. Indeed, since the adjoint string breaks into two gluelumps with mass  $\sim 2 $ GeV, one can estimate $r_{\max} \sim \frac{2M^{(2g)}}{\sigma_{\rm adj} } \approx \frac{4{\rm GeV}}{0.18 \frac94 {\rm GeV}^2} \cong 2$ fm.   A similar situation can occur in the $G_2$  group theory, and other special groups  \cite{26,27,28,29,30,31,32,33}.


In FCM the behavior of field correlator clearly agrees with that  found on the lattice \cite{38,39,40,41,42}, since both $D^E$ and $D^E_1$ contain perturbative terms $O\left( \frac{1}{x^4}\right)$ and $np$ terms $\exp (-\mu |x|)$, with $\mu \approx O(1$GeV), corresponding to the gluelump mass. Finally, one should address the important point of numerical compensation of  saturation terms $V_D^{\rm sat}(R)$ and $V_1^{\rm sat} (R)$ in (\ref{8}) and (\ref{11}), which otherwise would spoil the linearity of  confinement.
Indeed both potentials have opposite signs and similar magnitude, and one can easily check, that they compensate each other at the order of $O(10\%)$, when $M^{(1g)} \la M^{(2g)}$.

2. The accuracy of the Casimir scaling in QCD and SU(N) theory is associated with  the magnitude of higher terms $\frac{1}{n!} I_n$ in  the cluster expansion of $W(C)$, Eq. (\ref{2}). Indeed, using Eqs. (\ref{3}) and (\ref{6}) one can conclude that $\frac{I_4}{I_2} \la \frac{1}{20}$ and  the accuracy of the CS should be  around $O(5\%)$, which is supported by lattice data \cite{21,22}, in particular,  in \cite{22} the  accuracy of CS is around 5\%. One should stress  that this result is directly connected to the smallness of $\lambda,$ i.e. the large value of the gluelump mass as  compared to $\sqrt{\sigma}$. This can be explained as a high stochasticity of the vacuum, where the mean value of the field strength
$<F>= f $ satisfies the condition $f\lambda^2 \ll 1$.

At this point one can  associate the  quantity $f\lambda^2$ with the elementary flux on the surface of the Wilson loop and compare it  with the  corresponding  flux of  the instanton, $f\lambda^2 =2\pi$, while the flux of a magnetic  monopole, placed on the  surface, is $f \lambda^2_{mm} = \pi$ \cite{56}, which explains   intuitively   the  range of the magnetic monopole mechanism of confinement, since in this case  $ W(C) = \exp \sum^\infty_{n=1} \frac{(if\lambda^2_{mm})^n}{n!} \frac{RT_4}{\lambda^2} = \exp \left(\frac{RT_4}{\lambda^2} \exp (if\lambda^2) \right) \approx   \exp\left(-\frac{RT_4}{\lambda^2_{mm}} \right) , $ $\sigma_{mm} \sim (\lambda^2_{mm})^{-1} \sim 0.18$ GeV$^2$, $ \lambda_{mm}\sim 0.5$ fm.

However, in the FCM case $f\lambda^2 \sim \sigma\lambda^2 \sim \frac{\sigma}{(M_0^{(2g)})^2} \sim O(5\%)$ and hence one has the picture of stochastic small fluxes in QCD. This picture agrees very well with the lattice measurements in Ref.~ \cite{21,22}.

3. In FCM the flux distributions, given by (\ref{15}), (\ref{16}), describe the flux tubes of constant radius which are stabilized at large distance $R$ between charges, as shown in Figs. 6,7,8,9, taken from \cite{49}. One can see a good agreement with  lattice data of \cite{43}, where  the distance between charges $R$ belongs to the interval $0.76$ fm $\leq R \leq 1.33$ fm.

In this first step of flux tube theory, given by (\ref{15}), (\ref{16}) (see \cite{49} for details), one neglects the perturbative excitations of the string and the width of the flux tube stabilizes at large $R$. In the next  order one should take into account an additional gluon in the flux tube,   which corresponds to the hybrid state of $(QG\bar Q)$. The physics of this static hybrid and the corresponding eigenstates were given in \cite{55} in the framework of FCM. On the lattice these excited states were examined in \cite{53, 54}.

The formation of the flux tube is often  considered in analogy with the Abrikosov fluxes in superconductors and therefore it is one of the main  arguments in favour of the DGL picture of  confinement in QCD. As a consequence, one  tries to find the color magnetic monopole degrees of freedom in the QCD vacuum.

However, as shown in \cite{19, 49}, the flux tube picture occurs naturally in the  FCM, as it is demonstrated by Eqs. (\ref{15}), (\ref{16}).  Indeed, the correlation length $\lambda$ is provided by the gluelump mass $M^{(2g)}_0$, which is created by  confinement, i.e. by nonzero $\sigma$, and again $\sigma$ is given by the $2g$ gluelump Green's function $D^E(z)$. This mass creation fixed the size of  flux tube and its stability, and  leads to asymptotic verification of the dual London equation.

One may call it  the ``microscopic'' magnetic monopole mechanism, based on microscopic fluxes $f^2\lambda \ll 1$ in contrast to macroscopic DGL mechanism,  implying the existence of massive and large magnetic monopole with mass $m_\chi$ and dual gauge field with mass $m_B$, and their relation $k=\frac{m_\chi}{m_B}$, which are not predicted by theory.
It is very important that in FCM the mass creation process refers to the gluon,  connected to another gluon on the  given surface of the area law (in the parallel transporter), or to another propagating gluon in the $gg$ glueball, but never refers to one gluon separately. This is important for the problem of color Coulomb screening due to confinement , which is not still observed on the lattice \cite{22} and in the heavy quarkonium  spectrum \cite{57} \footnote{The highly excited states of charmonium and bottomonium may be  sensitive to the color Coulomb  screening, which strongly decreases the  dielectron width \cite{48}.} .

Indeed the total potential  between static charges both in FCM and  on the lattice  has the form
\be V_{\rm static}^{(j)} (R) = {\rm const}_{j} + V_{\rm Coul}^{(j)} (R) + \sigma^{(j)} R\label{17}\ee  with \be V_{\rm Coul}^{(j)} (R) = -C_2 (j) \frac{\alpha_s}{R} \label{18}\ee
with no appreciable screening mass $M_{\rm Coul}$ and  in FCM at zero temperature const$_j$ can be put equal to zero. At the same time in the field disrtibution of the flux tube the Coulomb field contribution to the total  colorelectric field $\veE \left( \frac{\veR}{2} \right)$ at the midpoint $\frac{\veR}{2}$ between static charges is equal to \cite{49}.
\be \veE^{(1)} \left( \frac{\veR}{2} \right)= 8C_2 (j) \frac{\alpha_s \veR}{R^3} \zeta    \left( \frac{ R}{2 \lambda_1} \right), ~~ \zeta (x) = (1+x) e^{-x},\label{19}\ee
which screens with the screening mass $M_0^{(1g)} = 1/2\lambda_1$, and this behavior is well supported by the lattice data \cite{43}.

Thus the FCM  theory explains both properties: 1) additivity of color Coulomb and confinement fields in the total $\veE$  and  2) the absence of screening in the $V_{\rm Coul} (R)$, while the screening is present in the flux  tube probes.

Indeed, the additivity is based on the additive form of the field correlator (\ref{4}), which for the total coloreletric field $\veE$ yields \cite{49}
\be E_i (\ver, \veR) = n_k \int^R_0 dl \int^\infty_{-\infty} dt \left(\delta_{ik}  D^E (z) +\frac12 \frac{\partial (z_i D_i^E (z)}{\partial z_k} \right), \label{20}\ee where $\ven = \veR/ R$. Calculating (\ref{20}) with $D_1$, given by the one-gluon gluelump Green's function with mass $M_0^{(2g)}= 1/\lambda_1$,  where one can approximate $\lambda_1 \approx \lambda$, one obtains the screening factor $\zeta \left( \frac{R}{2\lambda_1}\right)$ for large distances. This refers to the colorelectric field, measured as a probe in the Fig.4.

However, another result obtains if the static potential $V_{\rm Coul} (R)$ is calculated via $D_1^E(z)$ (see Appendix 2). One can  attribute this difference to different Wilson loop  constructions in these cases: it is of type of the Fig. 4 for the $\veE(\ver, \veR)$ in the probe plaquette and of the Fig.10 for the $V_{\rm Coul} (R)$. In last case the screening mass is  strongly suppressed due to small change in the area of the covering surface of Wilson loop, lifted by a propagating gluon.

An opposite situation occurs in the interaction of Wilson loops, studied within the FCM theory in \cite{47} and on the lattice in \cite{46}. A particular case of this analysis is the so-called disconnected probe of the field distributions of the static $Q\bar Q$ system, which was measured  in  \cite{12}.

Finally, the FCM analysis of flux tubes for three static charges in fundamental and adjoint irreducible representation (irrep)  leads to the pictures of 3 quarks with a central string junction in the first case and the triangular configuration in the second case,  as shown in Fig.11 taken from \cite{19}, in agreement with lattice and other data.

4. Till now we have discussed the case of static charges, when scalar or  vector type of the confining interaction $V_D(R)$ is not important. When going to the finite mass quarks, let us consider a light quark in the field of  an infinitely  heavy antiquark, where the Lorentz nature of $V_D(R)$ becomes crucial. Indeed, as shown in \cite{44,45}, in the case of the vector confinement the bound state spectrum of the Dirac equation does not exist and it excludes the possibility to use the  vector confinement of the Gribov-Zwanziger approach \cite{50} ( an exception takes place in the $d=2$ QCD, if some special transformation of the Bogolybov-Valatin type is done, see \cite{58}).

The scalar nature of the FCM confining interaction can be directly deduced from the form of $V_D$, expressed via $D^E(z)$, see  Appendix 2 . Then one can see the two-gluon-line exchange form for $V_D$ in contrast to the one-gluon-line form for $V_1^E$, which presupposes the scalar nature for $V_D$ and the vector nature for $V_1^E$ potential. An additional analysis, done in \cite{59, 60}, supports this conclusion.

5. In all previous analysis we have used the only parameter $\sigma$, while another parameter, $\lambda$, is expressed numerically via $1/\sqrt{\sigma}$, $ \lambda = \frac{1}{M_0^{(2g)}}\cong \frac{1}{4\sqrt{\sigma}}$. This situation was checked in \cite{34,35} by calculating the resulting $\alpha_s(M_0^{(2g)})$ and $\Lambda_{QCD}$ from the selfconsistency equation
$\sigma_f =\frac12 \int D^E (z) d^2z$ with $D^E(z)$, expressed via $\sigma$ and $\alpha_s$ as in  (\ref{8}), (\ref{9}). In this way one indeed has the only scale in the confinement mechanism (neglecting the quark mass modification of the confinement, which occurs in higher orders of $O(g^n)$).

7. The confinement mechanism described above is applicable both to QED and all $SU(N)$ theories, where one explicitly introduces the field correlators $D^E(z)$, $D_1^E(z)$, as in (\ref{4}). In the QED case, however, as shown in \cite{17}, one can apply  to the correlator $D_{i\alpha k\beta} (x,y)$ in (\ref{4}) the operator $\frac{\partial}{\partial x_\gamma} \varepsilon_{i\alpha \gamma\delta}$ and take into account the Abelian Bianchi  identities, $\partial_\alpha \tilde F_{\alpha\beta}=0$, which yields $D_{QED} (z) \equiv 0$, and hence no confinement. One can use the same technique of Gaussian correlators plus gluelump Greens function to other groups, like $G_2, F_4$ etc. and obtain confinement at intermediate distances, as it was observed on the  lattice \cite{26,27,28,29,30,31,32,33}. As it was discussed above, a screening of $V_D(R)$ due to string breaking may happen in the adjoint loops in $SU(N)$.

\section{The DGL  approach as the theory of confinement}

The DGL approach, suggested in \cite{2,3,4} and developed in numerous papers, is reviewed  in \cite{8,10,11}. In the DGL  the original Lagrangian can be written in the abelian Higgs form (the dual form of it is finally used).
$$ {L}=-\frac14 F^2_{\mu\nu} - |D_\mu \varphi|^2 - \frac{\lambda}{4} (|\varphi|^2 -\varphi^2_0), ~~ D_\mu = \partial_\mu -ie A_\mu$$which obeys the classical solutions -- the Abrikosov--Nielsen--Olesen (ANO) strings, and to describe confinement one needs a region of large $\lambda$, when one has a  condensate of electric charges $\varphi = \varphi_0$. It is clear,  that in this case the field $A_\mu$ acquires the mass $m^2_B = 2 e^2\varphi^2_0$, while the Higgs field has its own mass $m^2_\chi=2\lambda\varphi^2_0$. One can  easily follow the appearance of London's equations $\Delta \veB - m^2_0 \veB =0$, implying the stability of magnetic strings with the string tension $\sigma_{ANO}= \pi m^2_B \ln \frac{m_\chi}{m_B}$. Note the possible difficulty in detecting two  different mass scales: $m_\chi, m_B$ where $m_\chi$ should be much larger than $m_B$ in the proper dual Abrikosov scenario, whereas as we discussed above, one can see on the lattice the only mass  scale around 1 GeV. 
 Now we consider the properties 1-7, presented in the Introduction, with respect to the results of the DGL theory.

1. The linear behavior of confinement can be ensured by DGL mechanism with a good accuracy (see e.g. \cite{8,28} and refs. in \cite{11}) in limited region, which depends on $\lambda \sim \frac{1}{m_B}$, and the linear behavior is violated in the region $r\sim \lambda$.

 In DGL the behavior of the  quadratic field correlators, $G_2(x) = \lan F\Phi F\ran$,  as a function of $x$ was investigated in \cite{61*,10}, with the result: $G_2 (x) = \frac{c_1}{x^2}e^{-\mu x} + \frac{c_2\exp (-\mu|x|)}{x^4}$, which differs from the lattice data \cite{38,39,40,41,42}.

2. The Casimir scaling is strongly violated for fixed values of model parameters $m_\chi, m_B$ and $k=\frac{m_\chi}{m_B}$. Indeed, to reach an agreement with the CS values for higher representations  $J$ one needs large values of $k\ga 6$ changing with $J$ \cite{8,25}.

As we have discussed above in the previous section, the probable  reason for this behavior lies in the large values of elementary fluxes $f\lambda^2 =O(1)$, which bring into action the quartic and higher order correlators, $I_4\approx I_2$ according to Eq. (\ref{6}). Large values of $k$ imply the Abrikosov vortex mechanism of the second kind, however, the flux tube profiles   require lower $k$ values in the domain  of dual superconductors  of  the first kind \cite{43}.

3. The description  of the flux tubes in terms of the DGL theory is rather successful, as   shown in \cite{43}, however, to reproduce the actual change of the  flux tube profile, in \cite{43} it was used  the flux tube ansatz \cite{61}, based on the type I superconductor model, with three parameters dependent on the charge separation $R$. E.g. the parameter $k$ for $R=0.76$ fm, $R=0.95$ fm, and $R=1.33$ fm in Figs.6,7,8,9 should be chosen as $k=0.348, 0.170$, and 0.236, respectively.

At the same time, as shown in Figs.6,7,8,9, the FCM produces  distributions for all $R$ with the only parameter $\lambda=0.2 $ fm, which is connected to the  gluelump mass, calculated via string tension.

One can conclude, that DGL model corresponds to  general picture of flux  tubes, at it was expected, however, the microscopic structure of flux tubes is  not  yet described by a unique theory of the DGL type.

There are also attempts to describe the flux tube as the quantized Nambu-Goto string in the form of the Arvis potential \cite{62}  and the induced so-called L\"{u}scher term. However, the latter violates Casimir scaling \cite{21,22}. As it is known, the structure of string excitations is associated with an additional gluon degree of freedom, as found on the lattice \cite{53,54} and in the FCM \cite{55}, while  in the standard string theory \cite{61**,63*} the excitations have different structure.

Summarizing, one can conclude that in the DGL model the flux tube structure is resemblant to that obtained on the lattice for $SU(3)$ theory, but agrees with the latter qualitatively, since there was not found a unique set of parameters describing the lattice data.

4. It seems natural that in the DGL model one can obtain the scalar confinement, however, corresponding analysis is not known to the author.

5. The scales and self-consistency of the DGL model was not treated systematically. One clearly defines the scales of dual gauge field $m_B =1/\lambda$ and  the dual monopole field $m_\chi =1/\zeta$, and their ratio $k=\frac{m_\chi}{m_B}$,  but their connection to  the  only $SU(3)$ parameter, $\sigma$, also depends on other parameters, (dual ``Higgs'' coupling constant $\lambda_h$, and its vacuum  average $v$), so that as in \cite{8}, one obtains $\sigma_f=4\pi v^2$ in the Bogomol'nyi limit.
There are no examples of self-consistency checks, where all parameters are deduced from $\sigma$ or $\Lambda_{QCD}$.

6. Within the DGL model the Wilson loop-loop or string-string interaction was  investigated in the form of the disconnected probes \cite{12}. The point of the color  Coulomb non-screening  was not raised in the published literature, to the knowledge of the author.

7. It is clear that the DGL model can be formulated for any theory, where the gauge degrees of freedom are made massive using the dual Higgs field, taken from outside; therefore the main problem is to   identify both   gauge and dual Higgs d.o.f. This kind of separation and identification  is done  e.g. via the Abelian Projection Method \cite{13}, the center vortex model \cite{14},  thick vertex model \cite{15}, etc. These ideas can be equally well applied to other groups, like it was done in \cite{29} for the groups $G_2$ and $SU(N)$. It is difficult to judge whether the DGL explanation is successful unless one derives all parameters directly from the field Lagrangian itself and its renormalization constants.  Unfortunately, existing DGL applications are based on the parametrization of given theories in terms of assumed dominant d.o.f. and dominant structure.
\section{ The abelian projection and the center vortex model}

One of the most popular version of the confinement -- connected studies is the abelian projection  models.

Here one can use the maximally abelian gauge to separate the field configurations which are believed to contribute maximally to the phenomenon of confinement, see   reviews \cite{10,11,13*,14}. Ideologically this  direction is connected to the idea of monopole dominance, since the necessary gauge transformations might include singular gauge configurations  reminiscent of magnetic monopoles.  The main idea of the  abelian projection is to extract from the nonabelian field   monopole d.o.f. and to this end one can write in maximally abelian gauge (MAG) in the SU(2) case  the gauge-transformed plaquette $U_{\mu\nu}$ as $U_{\mu\nu} = \sum_i \exp (i \theta^{(i)}_{\mu\nu} \sigma_3)$, with the separation $\theta_{\mu\nu} = \bar\theta_{\mu\nu} + 2 \pi n_{\mu\nu}$, where $-\pi\la \bar \theta_{\mu\nu} \la \pi$ is   the abelian part and $n_{\mu\nu}$ is the ``monopole part''. The resulting contribution of both parts to the static potential is shown in Fig.12 \cite{67*}, where one can see the dominance of the ``monopole part'' in the string tension and  the nonconfining abelian ``photon'' contribution. In the lattice studies  \cite{67**, 70*}  are presented accurate calculations of static quark potential in the maximal abelian gauge (MAG) in comparison to the  exact lattice data for $Q\bar Q$ and $3Q$ configurations. One can see in \cite{67**} a very good agreement between MAG and exact values of $\sigma$ with accuracy of the order of 5\%. At the same  time the color Coulomb part of the interaction disappears in the MAG version.

How one can understand these results from the point of view of FCM? To this end one must remember that the field correlators $\lan FF\ran$, responsible for confinement, are produced by the gluelump Green's function, where the color links can be considered as diagonal in the color space in the lowest (nonperturbative) approximation, whereas the color change is provided by the perturbative vertex $gf^{abc} A_\mu^b A_\nu^c$. Hence one can expect that the color diagonalization would keep all the confinement effects up to $O\left(\frac{\alpha_s}{\pi} \right)$ and the color exchange potential, generated by perturbative vertices, should be absent in the color diagonal approximation.

In this way the results of the abelian projection method can be connected to and explained by the FCM. 

There appears another question: what are physical contents of the abelian projection method (APM)? It is clear that for any physical mechanism of  confinement, e.g. for  the  FCM, the resulting field distributions can be analysed via the APM, and the only way to explain the confining result is the presence of the $n_{\mu\nu}$ piece in the  plaquette, which however has nothing to do with the real monopole admixture, but rather with the fact, that purely abelian d.o.f. do not ensure confinement, and one needs the ``elementary monopole d.o.f.'' at each point.

The center vortex model (CVM) \cite{14} and its  extension  -- the thick center  vortex model (TCVM) \cite{15} have attracted a serious attention during  last 20  years (see \cite{26,27, 28, 29, 30, 31} and refs. therein). The  main idea of these models, as well as in the DGL approach, is to represent the QCD vacuum as  stochastic ensemble of color magnetic fluxes in terms of the nontrivial center elements of the gauge group, and  these vortices are thickened  in  the TCVM, \cite{15} which  allowed to build up the linear potential for all representations at the intermediate distances.
As one can see, fundamentally the CVM (TCVM) is based on the notion of the stochasticity  of field fluxes connected to center vortices  in their total number and orientation. From this point of view there is a  similarity between CVM (TCVM) and the FCM approach,  since in FCM one has a stochastic ensemble of  field fluxes $\hat F_{\mu\nu} (z) d  \sigma_{\mu\nu} (z)$ inside the Wilson loop, where $\hat F_{\mu\nu} (z)$ is the  gauge covariant field strength defined in Eq. (\ref{2}). 
The stochasticity of the ensemble  $\left\{ \hat F_{\mu\nu} (z_n), n=1,2,...\right\}$ on the plane of  the Wilson loop follows from the short  correlation length  $\lambda =   \frac{1}{M_{Glp}} \la 0.2$ fm, and as was discussed above (Eq. (\ref{6})), the quartic and higher correlators violate the Casimir scaling  by less than  5\%. Note, that the topological or group  structural  properties are not necessary for the  resulting  confinement. 

A more detailed group theoretical structure is assigned to the  independent field fluxes in CVM (TCVM),  where each  flux is  connected to a center element.

A detailed numerical analysis of the SU(2)  Yang-Mills theory  in the framework of TCVM was performed in \cite{15}, and the linear potentials  of  static sources have been obtained at  intermediate distances. As one can see  in Figs.1,2 of \cite{15}  the linearity of potentials,  especially for hiher representations is indeed achieved in the intermediate region, however there is a problem with the Casimir scaling in this linear region, which is violated for higher representation up to 25\%, in contrast to the accurate $(< 5\%)$   scaling in direct lattice calculations. 

Another interesting problem in this approach is the significance of the center of group, which can be trivial as in the case of the $G(2)$ group, while SU(2) and SU(3) subgroups  are present in $G(2)$. This topic was  discussed  in \cite{28,31,59'}. It was found in particular in \cite{31} that in $G(2)$ there are two linear regimes at lower $R/a \approx 5$ and $\frac{R}{a} \sim 25$, and for  the fist one  the Casimir scaling is qualitatively valid.

It is clear, that the FCM method can be   directly applied  to the simple group theories, e.g. to the $G(2)$ Yang-Mills model and  these results can be compared to those in \cite{28}  and \cite{31},  which can  establish a link between the two approaches, in particular it wood be interesting to connect vortex probabilities  with  the corresponding field correlators.

Here the criterion of the linear confinement between static charges in the $G(2)$ (or any other field theory without quarks) is  that all field correlators $\lan \hat F_a (x_1)  \hat F(x_2)... \hat F_c (x_n)\ran$  should have  exponential asymptotics $exp (- m_{ij} |\vex_i- \vex_j|,...)$ with $m_{ij}> 0$,  and the  linear confinement occurs in FCM for $R> 1/m_{ij}$.

As it is, the comparison between the two methods can reveal  additional properties of the confinement phenomenon. 
\section{Scalar or vector confinement  in QCD}

Our  consideration   above has to  do with  static potentials, where the scalar or vector character of interaction is not important (as will be also seen  below). Here we study the  case of quarks of any  mass and start  with the case of a quark of an  arbitrary mass $m$ in the field of a static charge. Following \cite{44,45} we write the Dirac   Hamiltonian
\be H =\veal \vep + \beta m + \beta  U(r) +V(r), ~~ H\psi = E\psi\label{27}\ee
where $U$ and $V$ are scalar and vector potentials respectively. In the  standard bispinor  formalism   $\Psi =\frac{1}{r} \left(\begin{array}{l} G(r) \Omega\\iF(r) \Omega'\end{array}\right)$, one arrives at the system of equations
\be \frac{dG}{dr} +\frac{\kappa}{r} G - (E+m +U-V) F=0\label{28}\ee

\be \frac{dF}{dr} +\frac{\kappa}{r} F + (E-m -U-V) G=0\label{29}\ee
Assuming  $U=\sigma_sr,  ~V=\sigma_vr,$,   we shall consider three possibilities
(i) $U\neq 0, V=0$, (ii) $U=0, V\neq 0$, (iii) both $U, V\neq 0$.

Introducing notations $x= \sqrt{\sigma_s} r, \varepsilon = E/\sqrt{\sigma_s}$, in the case (i) one obtains solutions  with the asymptotic \be G,F \sim \exp \left( - \frac12 (x^2 + bx )\right), ~~ G^{''} - x^2 G \approx 0,\label{30}\ee implying a reasonable bound state problem even for $m\to 0$. In the  case (ii) replacing $\sigma_s \to \sigma_v$, one arrives at the equation

\be G^{''} + x^2 G =0, ~~ G\sim \exp \left( \frac{ix^2}{2} \right).\label{31}\ee

Thus one cannot have bound states in the vector potential.

In the  case (iii) with  definition $\sigma_s=c_u \sigma_0, \sigma_v =c_v \sigma_0, x=\sqrt{\sigma_0} r$, one  finally obtains asymptotically at $x\to \infty$,\be G^{''} - \left[ \left( c_u x + \frac{m}{\sqrt{\sigma}}\right)^2 -  c_v x^2 \right] G\approx 0.\label{32}\ee

From (\ref{32}) one can deduce that 1) the necessary condition for the bound state spectrum is $c_u>c_v$, and 2) for $m\to \infty$ the bound states exist for  any type of confinement.

Comparing this situation with the Gribov-Zwanziger model of confinement \cite{50} one can conclude that the linear vector confinement of this model is not  compatible with QCD, unless some additional vacuum transformation of the Bogolybov-Valatin type is possible, producing finally the scalar confinement,   as it  happens in the $D=2$ QCD \cite{58}.

\section{Additional tests  and  consequences of the  confinement mechanism}

When applying confinement mechanism to real hadron physics, one meets with numerous applications, which serve as a serious test of its nature. Below we shortly discuss several important applications, which should be present in any approach to confinement.
\begin{enumerate}

 \item QCD string and Regge trajectories;

 \item confinement in the  fast moving hadrons;

 \item the role of confinement in the  IR regime and  the  convergence of the perturbative  series;

 \item confinement and chiral symmetry breaking;

 \item  confinement and deconfinement at finite temperature.

 \end{enumerate}
 Below we  shortly consider  all these topics comparing results of  confinement mechanisms with real phenomena in hadron physics.

 \subsection{ QCD string and Regge trajectories}

 There are two main approaches to the definition of  string  spectrum, in the first  one considers the QCD  string, created by confinement, as an example of the string theory \cite{61**,63*} renormalizable in higher dimensions,  with its characteristic  spectrum of excitations, where each point of string is a dynamical variable and the spectrum is  the collective excitation. This type of dynamics was used to calculate contribution of excited string states to the so-called Arvis potential \cite{62}  and  Nambu-Goto type strings \cite{61**,63*}.

 Another approach to the QCD string and the QCD spectrum in general follows from the FCM and can be called the FCM string, where the motion of string is  defined by its boundary, i.e. by the ends of the string in the mesons, baryons, and glueballs, $\bar qq, 3q, 3g$ and will be called configuration  $A$; in other case there are extra gluons ``string on the string in the  excited states ", i.e. $\bar q gq, \bar qggq, ...$ i.e. hybrids, which will be called configuration $B$.

 In the FCM string the parts of the string between the ends or  internal gluons can be considered as inert and their motion is defined by the end   points. This is clearly  demonstrated by the  Nambu-Goto form of the action,  exploited in the  basic papers of this direction for Regge trajectories \cite{63,64,65}.

 Using the relativistic Hamiltonian dynamics \cite{66}, one obtains the well-known form of the QCD string Hamiltonian \cite{63,64,65}, describing orbital Regge trajectories of mesons \cite{64}, radial meson Regge trajectories \cite{65}, Regge trajectories for baryons \cite{67}. There it was demonstrated that this theory works very well for all orbital (proper) Regge trajectories, while for radial Regge trajectories one needs to take into account the  flattening of confining potential at large distances -- as it is known in QCD with  light quarks.   The  similar trajectories for glueballs are calculated in \cite{68,69}, where also  the connection with the pomeron trajectory was studied \cite{69}.

 The QCD string excitations as  hybrid states,  introduced in FCM \cite{55}, have been compared there  with lattice data  \cite{53}, displaying  a reasonable agreement, the same can  be said about Regge trajectories in  \cite{63,64,65}, where comparison was done with  experimental hadron masses. Summarizing this part, one can   say that the QCD string picture based on FCM, where the string pieces are  treated  as inert objects without dynamical d.o.f., is in  reasonable agreement with experiment.

 \subsection{Confinement in the fast moving hadrons}

 It is widely known that in the high-energy collisions fast partons are considered as free particles, not  subject to any confinement interaction. This seemingly universal proposal is in  contradiction with the  relativistically invariant picture, where the transverse d.o.f. can be independent of the (fast) motion of the object.

 Indeed, recently in \cite{77} the author has  constructed the relativistical invariant Hamiltonian of a fast moving hadron with the  resulting solution for the boosted wave function, where the transverse coordinate (momentum) dependence was independent of the  motion and coincided with behavior of the wave function in the rest frame. E.g. for the  $S$-wave $q\bar q$ hadron the boosted w.f. can be  written as \cite{77}
 \be \varphi (p_\bot, x) =\varphi \left( \sqrt{ p^2_\bot + \left(x_1 - \frac12\right)^2 M_0^2} \right),\label{38}\ee
 where $M_0$ is the rest mass of the hadron.

 As a result, one obtains the valence $pdf$ (the parton distribution function) in the hadron
 \be D^q_n (x,p_\bot) = \frac{M^2_0}{(2\pi)^3} |\varphi (p_\bot, x) |^2.\label{39}\ee

 In the total $pdf$ one should take into account all other Fock components, which in the  standard formalism \cite{78} are given by  free sea quarks etc.,  while in this ``boosted confinement'' picture they are represented by fast hadron and  hybrid states. The corresponding picture was developed in \cite{79} and is in good agreement with DIS  experiment.

 As  an additional point  the account of the  contracted wave functions (\ref{38}) in the process of $e^+e^-\to hh$  allows to predict the asymptotics of form factors of mesons and baryons  (the famous ``power law''  but without gluon  exchanges)

 \be F_h (Q_0) \sim \left( \frac{M_0^2}{Q_0^2}\right)^{n_h}, n_M =1,~~ n_B=2\label{40}\ee where $Q_0= \sqrt{M^2_0+\vep^2}$.

 This topic is of crucial importance for our understanding of high energy (HE) processes and is to be developed further.

 In our confinement  treatment one can conclude that in the fast moving object the transverse part of confinement  is kept intact, so that in all HE collisions, where deconfinement does not occur, all processes are proceeding with  confined quarks and gluons in the form of (also) highly excited hadrons, i.e. multihybrids.

 Thus confinement strongly affects the standard picture \cite{78} also in the HE process. For   considerations,   concerning the role of confinement in  HE processes and  the HE momentum sum rule, see \cite{80}.

 \subsection{Confinement, IR divergence, and the  convergence of  perturbative series}

 The perturbation series in QCD is subject to  serious internal difficulties, as it is known in the textbooks \cite{82}, and we will consider those in the following order.

 a) The factorial divergence due to IR renormalons \cite{83}.

 b) Landau ghost problem for $\alpha_s$ in the IR region \cite{84}.

 c) IR divergencies in QCD, being treated as in QED, neglecting confinement \cite{78}.

 d) The ``Euclidean character'' of the perturbation theory (PTh) due to absence of hadron thresholds in the analytic dependence $\alpha_s (Q)$ etc.

 To introduce confinement in the PTh one can use the Background Perturbation Theory (BPTh), as formulated in \cite{85} and developed further in \cite{86,87} with the help of confinement.

 As a result one arrives at the PTh, where  for all gauge invariant amplitudes, in the higher orders containing closed loops, one  takes into account    the area law due to confinement. This property strongly changes the behaviour at low momenta and e.g. for the IR renormalon series of amplitudes one arrives at the resulting sum, which does not contain factorial growth at all \cite{88}, thus solving the long-standing IR renormalon problem a).

 b) In a similar way, taking into account the confining loops in the renormalization of $\alpha_s$ in the next orders, one arrives at the IR finite form of $\alpha_s$ \cite{84}, namely, for the one loop one obtains
 \be \alpha_s (Q) = \frac{4\pi}{\beta_0} ln \left ( \frac{Q^2+M^2_B}{\Lambda^2_{QCD}}\right), \label{41}\ee
 where $M^2_B = 2\pi\sigma, M_B \cong 1$ GeV. This result is in a good agreement both with $\alpha_s (Q^2)$  for $Q^2\ga 3$~GeV$^2$, as well as with lattice  expectations \cite{89,90,91}.

 c) In the HE evolution theory there is a standard agreement that the IR divergence of the QCD amplitudes  can be treated similarly to QED, i.e. introducing the gluon mass or the lower limit of the gluon energy $w_{\min}$ to make the divergent amplitudes finite \cite{78}.

 However confinement precludes the free motion of quark and gluons, placing those on the confining film of the Wilson loop. As a consequence any amplitude of the QCD PTh is  IR finite at nonzero $\sigma$ (for examples of IR regularization due to confinement see \cite{92} and appendix therein). This fact must modify the standard PTh and $pdf$ evolution equations, see \cite{79} for  recent results.

 d) The inclusion of confinement in PTh  automatically introduces hadron spectra and hadron thresholds $M_{th}$, and usually in $\alpha_s(Q)$ the matching procedure at the current quark masses is used \cite{78,82}. Note, that $\sigma$ in $SU(N)$ plays the same role as $\Lambda_{QCD}$ and one can be expressed via another, as it was shown in \cite{34}. The same role of $\sigma$ retains in QCD  with quarks, so that the QCD PTh with confinement contains the same number of RG parameters as the standard PTh, namely $\sigma$ and quark masses $m_q$.

 \subsection{Confinement and chiral symmetry breaking}

 Confinement and the CSB  are internally connected, since  the CSB is known only in the  systems, where confinement is operating. Moreover, at the  growing temperature $T$ the chiral  condensate is vanishing just in the region, where the  presence of confinement cannot be substantiated.

 As was  discussed in section 5, the  resulting confinement for light quarks  should be of scalar character (it was argued in \cite{44,45} ).  As a result the presence  of the scalar term $\bar q M q$ in the effective Lagrangian signals about the CSB.  The  whole point now is to provide the formalism which ensures both confinement and CSB, and yields all  known relations (e.g. GMOR), derived  earlier without connection to  confinement.

 This formalism was created in \cite{93} and generalized in \cite{94}, where the effective Lagrangian for quarks with account of  confinement and CSB is written in the form
 \be L_{eff} =-N_c tr log (i\hat\partial + \hat m + M(\vex) \hat U), ~~ \hat U = \exp (i\hat \phi \gamma_5). \label{42}\ee

 Here $\hat U$ contains standard Goldstone pseudoscalar mesons, and $M(x) \approx \sigma |\vex|$ takes into account confinement of the quark with the antiquark (it is convenient to associate $\vex  =0$ with the midpoint between $q$ and $\bar q$). It is also important that  at the vertex  of the $q\bar q$ Green's function,  $|\vex| =0$  however at $x\la \lambda$ the simple linear behavior is replaced by a more  complicated form,  to  simplify the matter we impose the  boundary condition  $M(0) \to M(\lambda) = \sigma \lambda \simeq 0.15$ GeV.

 The Lagrangian (\ref{42}) allows to obtain all  known CSB relations \cite{93,94}, where $M(0)$ does not enter, but in addition many new relations, e.g. the Effective Chiral  Lagrangian (ECL) was derived with known coefficients not only in $O(p^2)$ order, but also in $O(p^4), O(p^6)$ \cite{94} . Moreover, the quark coupling constants of $\pi, K -- f_{\pi}$ and $f_K$ have been calculated in good agreement with lattice and experimental data \cite{95}. The most important point here is that the new chiral -confinement (CC) Lagrangian takes into account both $q,\bar q$,  and the chiral $(\pi,\pi\pi, K, K\bar K$,...)  degrees of freedom. This is important in the case of  external fields, where   e.g. magnetic fields  (MF)  act directly on the quark  d.o.f. and indirectly on the chiral d.o.f. (e.g. on $\pi^0, 2\pi^0..$). This important check was done  in the MF  dependence of quark condensate, $ \lan \bar q  q (B) \ran $ in  \cite{96}. Both results are in good agreement with lattice calculations, which, however, differ from  earlier pure chiral d.o.f. results \cite{99*}.

 It is important that in all these  cases confinement enters  only via string tension $\sigma$  and  sometimes via $M(0) =\sigma \lambda$, where $\lambda$ is the  same as in the gluelump mass, $\lambda = \frac{1}{M_{glp}} \approx 0.2$ fm.

 In calculations  of the $\pi$ and $K$ masses in MF  via $\sigma$ \cite{97}, one finds again the dominance of the $q\bar q$ over purely chiral d.o.f.

 Finally, it is very interesting what happens with CSB with growing quark masses and  at  which value of $m_q$ the chiral properties are lost. This analysis was done  in the FCM  \cite{98} and compared with lattice and other data. As one can see in \cite{98},  the spectra of PS mesons approach the non-CSB form at  $m_q \ga 150$ MeV, in agreement with lattice data \cite{99}.

 \subsection{Confinement and deconfinement at growing $T$}

 Both $SU(N)$ and QCD with  $n_f>0$ are subject to   the deconfinement process at (or around) some transition temperature $T_c$, and at higher $T$ on the lattice one observes the quark-gluon medium, called the  quark-gluon plasma (qgp). While in SU(3) this is the weak first order transition \cite{100}, in the case of QCD, $n_f=3$ the   thermodynamic mechanism is not yet determined. These transitions can be understood  qualitatively from the principle of the minimal free energy (maximum pressure), if one  neglect artificial Hagedorn states.

 The interesting  point  is that confinement is $T$ - dependent and in SU(3) $\sigma$  decreases before the transition, as  shown in \cite{101}, which helps to  describe  well the  whole $T$ dependence \cite{100}. The  same happens in QCD with  $n_f=3$,  and  here the visual phenomenon is  the  disappearance of  the chiral condensate $\lan \bar q q(T) \ran $  at $T\sim 0.16$ GeV \cite{102}, since in the FCM  at small $m_q $,  $\lan \bar q q (T) \ran \sim ( \sigma (T) )^{3/2}.$

 Since FCM is $O(4)$ invariant, it contains both colorelectric (CE) and colormagnetic (CM)  correlators, which imply CE and CM confinement (CMC). At $T=0$ both CE and CM correlators  coincide, while with growing $T$ the correlators behave in a different way.

The  effect of CMC  is  dominant for the  gluon plasma in SU(3) and  for the $qgp$, as shown in \cite{100, 106},  where the CM dynamics was explicitly formulated and the results compared with lattice data. One can expect that CMC is also  important for $q \bar q$ and $gg$ correlations in the deconfined phase.

 The CMC plays an especially important role in the  temperature perturbation theory, where it prevents IR divergencies and allows to solve the Linde problems, by summing a  converging  infinite set of  finite perturbative  diagrams \cite{107}.

 Thus the deconfinement  process helps to understand confinement and its structure in detail, and  this analysis is becoming more and more informative.

\section{Conclusions}

The whole discussion above is aimed at the understanding confinement not  as a special interesting phenomenon, but rather    stressing the point that in QCD confinement is a central property of the whole physical world, which establishes its existence. Indeed, confinement via the baryon masses creates 99\% of the visible energy in the Universe.
Therefore all properties of confinement, listed as 1)-7) in Introduction and all five consequences in section 7, are intimately connected to each other.

We have shown above that the FCM satisfies all these criteria and is  intimately  connected to the bilocal field correlator, which provides confinement (the area law of Wilson loops) yielding mass to the  gluons inside a hadron. In  this respect FCM is similar to the DGL approach, but the FCM mechanism of the gluon mass creation is different. Indeed, for FCM  the quadratic correlator is dominant and this  dominance is  self-consistent and is supported by the Casimir scaling, while in DGL the  confinement mechanism evidently includes higher order correlators. It is interesting to search for the effects, connected to the quartic and higher correlators, possibly associated with the DGL  configurations and justifying its presence (at least partial) in the  QCD vacuum.

The simplicity of the FCM confinement and its capability to satisfy all criteria, makes it  reasonable theory, which can be successfully used in all nine directions, as it was demonstrated above.

With all that, there are still unresolved issues of confinement theory: 1) the mechanism of  deconfining temperature transition in full QCD, 2) the temperature variation of field correlators, yielding the  decreasing string tension $\sigma(T)$, 3) the role of higher correlators in QCD. The latter can be illustrated by the relation obtained in \cite{17,57}
\be
\frac{dD^E(z)}{dz^2} = \frac{g}{8} f^{abc} \lan F^a_{\alpha\beta} (0) F^b_{\beta \gamma} (0) F_{\gamma \alpha}^c(0)\ran .\label{45}\ee

These topics are presently studied in the framework of the FCM approach.

The author is grateful to A.M.Badalian for helpful discussions. This work was done in the framework of the scientific project, supported by the Russian Scientific Fund, grant No 16-12-10414.

\setcounter{equation}{0}
\renewcommand{\theequation}{A1.\arabic{equation}}

\hfill {\it\large   Appendix  1}\\

\Large{\centerline{\it            The contribution of  the correlator $\llan \hat{F} \hat{F}\hat{F}\hat{F}\rran$}}
\normalsize
 \vspace{1cm}

 The quartic correlator is equal to $\frac{I_4}{4!}$, where \be I_4 = g^4 \int_S d\sigma (1) \int_S   d\sigma (2)\int_S d\sigma (3)\int_S d\sigma (4)\llan \hat{F}(1) \hat{F}(2)\hat{F}(3)\hat{F}(4)\rran
.\label{A1}\ee

Since the correlator is selfconnected, i.e.  it does not depend on position for large area $S$, one can separate one integral, say $\int d\sigma (1) = area (S) =RT_4$, and  estimate the rest  as follows. The typical construction for the 4- point connected correlator  $G(1,2,3,4) \equiv \llan \hat{F}(1) \hat{F}(2)\hat{F}(3)\hat{F}(4)\rran$ with $2g$ vertices $g[A(i) A(i)]$ at each point $i$, is made with $1g$ propagators  connecting neighboring points and the   covering film over all construction,  so that an upper limit is
$$ G(1,2,3,4) \la g^4 \lan G^{(1g)} (1,2)G^{(1g)} (2,3)   G^{(1g)} (3,4) G^{(1g)} (4,1)\ran ,$$
where  $  G^{(1g)}_{\mu\nu}  (x,y)= \lan Tr_a A_\mu (x) \Phi (x,y) A_\nu (y) \ran$. Using the asymptotics of $G^{(1g)}$ in \cite{13}
\be  G^{(1g)} (x,y) \sim (N^2_c -1) N_c \frac{\sigma_{\rm adj}}{4\pi} e^{-M^{(1g)} |x-y|}= ae^{-M^{(ig)} |x-y|},\label{A2}\ee
one arrives at the estimate
\be I_4 = {\rm area} (S) \int d^2 x d^2 y d^2 z a^4 e^{-M^{(1g)} \sum_{ij} r_{ij}} \label{A3}\ee
with $a=\frac{gN_c (n_c^2-1) \sigma_{\rm adj}}{4\pi}$.

The  integral yields the  factor $\frac{\rm const}{(M^{1g})^6}$, with ${\rm const} \approx O(1)$, so that the final estimate is
\be I_4 = {\rm area} (S) \left( \frac{gN_c (n_c^2-1) }{4\pi}\right)^4\sigma_{\rm adj}
\frac{\sigma^3_{\rm adj}}{(M^{(1g)})} \approx \sigma_{\rm adj}{\rm area} (S) (\sigma\lambda^2)^3.
 \label{A4}\ee

 Since $\sigma \lambda^2 \approx 0.05$, one can see a strong suppression factor, ensuring the Casimir scaling at the order $O(F^4)$. \\


 \setcounter{equation}{0}
\renewcommand{\theequation}{A2.\arabic{equation}}

\hfill {\it \large  Appendix  2}

\centerline{\it\large }

 \vspace{1cm}

 We start with the standard definition of quadratic correlator \cite{13}
 $$ D_{\mu\nu,\lambda\sigma} (x,0) = \frac{g^2}{N_c} \lan tr F_{\mu\nu} (x)
 \Phi (x,0) F_{\lambda\sigma} (0)\ran =$$
 \be = (\delta_{\mu\lambda} \delta_{\nu\sigma} -\delta_{\mu\sigma} \delta_{\nu\lambda})  D(x) +\frac12 \left[ \frac{\partial}{\partial x_\mu} (x_{\lambda} \delta_{\nu\sigma}-x_\sigma \delta_{\nu\lambda} ) +(\mu\lambda \to \nu\sigma)\right]D_1(x).\label{a.1}\ee
  To express static potential $V(R)$ via $F_{i4}$ one can use the representation
  $$ W(R,T_4) =\exp \left( - \frac{g^2 }{2}\int  \lan F \Phi F\ran d\sigma d\sigma'\right)=\exp(-\hat V dt_4)$$ and  express the correlator $\lan F(x) F(y)\ran \sim D$  as in (\ref{a.1}). As a result one obtains  $$ \int \hat Vdt_4 = \frac12 \int D_{14,14} (u-v) d^2u d^2v=$$
 $$=\frac12 \int d\left( \frac{ u_4+v_4 }{2}\right)  d(u_4-v_4) d\left( \frac{ u_1+v_1 }{2}\right)  d(u_1-v_1)D_{14,14} (u-v) =$$
\be =2 \int dt_4 \int^\infty_0 d\nu \int^R_0 d\eta(R-\eta) (D(\nu,\eta) + \frac12 \frac{d}{d\eta} (\eta D_1 (\nu,\eta))\label{a.2}\ee
where $t_4 =\frac{u_4+v_4}{2}, ~\nu=|u_4-v_4|, ~\eta=|u_1-v_1|$, one finally obtains
\be  V =V_D (R) + V_1(R), ~~ V_D =2 \int^\infty_0 d\nu \int^R_0 d\eta (R-\eta) D(\nu, \eta)\label{a.3}\ee
\be V_D(R) = V_{\rm lin} (R)+ V_D^{(sat)} (R)\label{a.4}\ee
 and consider the colorelectric correlator $D^E_1 (x)$ putting $\nu=\sigma=4$, which produces the potential $V_1(r)$ .
 \be V_1(r) = \int^r_0 \lambda d\lambda \int^\infty_0 d\tau D^E_1 (\sqrt{\lambda^2 +\tau^2}).\label{a.5}\ee

  As  a new step one must express $D_1(x)$ via the gluelump Green's function to  the lowest order in  background perturbation theory. To this end one can  extract from $F_{\mu\nu} (x) = \partial_\mu A_\nu -\partial_\mu A_\mu - ig [A_\mu, A_\nu]$, the part with  derivatives, which contributes  to $D_1$
  \be  D_{1~ \mu\nu, \lambda\sigma}^{(0)} (x,y) =
  \frac{g^2}{2N_c^2}
   \left\{ \frac{\partial}{\partial x_\mu} \frac{\partial}{\partial y_\lambda} \lan  tr_a A_\nu (x) \Phi (x,y) A_\sigma (y) \ran +{\rm perm}\right\} \label{a.6}\ee
  and denoting  as $G^{(1g)}$ the  structure in the angular brackets, one obtains for $\mu=\lambda=4$.

  \be  \frac{\partial}{\partial x_4} \frac{\partial}{\partial y_4} G_{\nu\sigma}^{(1g)} (x-y)=\frac{\partial}{\partial x_4} (x_4-y_4) D_1(x-y) \delta_{ \nu\sigma}+{\rm perm}. \label{a.7}\ee

  As a result one obtains that $G_{\nu\sigma}^{(1g)} (z)= \delta_{\nu\sigma}  G ^{(1g)} (z)$  and $D_1^E$ is connected to  $G^{(1g)}$  as
 \be D_1^E (x) = - \frac{2g^2}{N^2_c} \frac{dG^{(1g)}}{dx^2}.\label{a.8}\ee
 Here $G^{(1g)} (x-y)$ is the Green's function of the gluon gluelump, i.e. the gauge invariant combination of the gluon propagator augmented with the parallel transporter $\Phi(x,y)$, as shown in Fig.1

 This function and the  corresponding gluelump spectrum was found   analytically \cite{36}, being  in a good agreement  with the lattice  data\cite{37}.

 Inserting (\ref{3}) into (\ref{2}), one obtains the relation between $V_1(r)$ and $G^{(1g)}$, namely
 \be  V_1(r) =- \frac{g^2}{N_c^2} \int^\infty_0 d\tau (G^{(1g)} (\sqrt{r^2 +\tau^2})-G^{(1g)} ( \tau  )). \label{a.9}\ee
 Since $G^{(1g)} (x\to \infty )\to 0$ one obtains
 \be  V_1(\infty) = \frac{g^2}{N_c^2} \int^\infty_0 d\tau  G^{(1g)} ( \tau  ).  \label{a.10}\ee
 \be V_1(r) = V_{\rm coul} + V_1^{(\rm sat)}\label{a.11}\ee

  Therefore to define properly the perturbative Coulomb interaction one can write,
  \be V_1^{\rm (pert)} (r)= \frac{8\alpha_s}{3\pi} \int^\infty_0 d\nu\left( \frac{1}{\nu^2}-\frac{1}{\nu^2+r^2}\right) = V_1^{\rm (pert)} (\infty)-\frac{4\alpha_s}{3r}\label{a.12}\ee
 and to  renormalize (\ref{15}) one can put $V_1^{\rm (pert)} (\infty)=0$.

In a similar way one introduces the two-gluon gluelump, as it was done in \cite{34,35,36}


 \be D_{ik lm}(x,y) = D^{(0)}_{ik,lm} + D^{(1)}_{ik,lm}+
D^{(2)}_{ik,lm}.\label{a.13}\ee

 \be D^{(2)}_{ik, lm} (x,y) =- \frac{g^4}{2N_c^2}\lan {\rm tr}_a ([a_i,
 a_k]\hat \Phi (x,y) [a_l,a_m])\ran.\label{a.14}\ee

   \be [a_{i}, a_k] = i  a^a_i a^b_k f^{abc}
T^c\label{a.15}\ee   \be
G_{ik,lm} ={\rm tr}_a\lan f^{abc} f^{def} a_i^a(x) a_k^b(x)
T^c\hat \Phi(x,y) T^fa^d_l a^e_m\ran.\label{a.16}\ee  \be G_{ik, lm}(x,y) = N^2_c (N^2_c-1)
(\delta_{il} \delta_{km}-\delta_{im} \delta_{kl}) G^{(2\rm {
gl})}(x,y),\label{a.17}\ee

 \be D(x-y)
=\frac{g^4 (N_c^2-1)}{2} G^{(2{\rm gl})} (x,y).\label{a.18}\ee
\be G^{(2{\rm gl})(0)} (x,y) = \frac{1}{(4\pi^2(x-y)^2)^2} +
O\frac{(\alpha_s\ln (x-y))}{(x-y)^4},\label{a.19}\ee
\be\sigma=\frac12 \int d^2 x (D(x)+{\rm
higher~correlators}).\label{a.20}\ee

\be G^{({\rm 2gl})} (T) =\sum |\Psi_n^{({\rm 2gl})} (0)|^2
e^{-M_n^{({\rm 2gl})} T}.\label{a.21}\ee\\

\setcounter{equation}{0}
\renewcommand{\theequation}{A3.\arabic{equation}}

\hfill {\it \large  Appendix  3}

\Large{\centerline{\it
 Selfconsistency  of $\sigma_f$}}
\normalsize
 \vspace{1cm}

The behavior of $G^{(2g)}(z), D^E(z)$
 for $z\to 0$ was analyzed in \cite{34,35}, where it was shown  that at $z\ll \lambda$, $D^E_{np} (z) \equiv D^E(z)$, after subtraction of the   divergent term, $G^{2g)}(z) =\frac{\rm const}{z^4}$

 $$ G^{2g)}(z) = G^{2g)}_{\rm pert} (z) +
 G^{2g)}_{np}(z)$$

$$G^{2g)}_{\rm pert} (z)= \frac{1}{(4\pi^2z^2)^2} +O\left(\frac{\alpha_s ln z}{z^4}\right)$$ has a mimimum at $z=0$,  namely $$D^E(0) \cong \frac{N_c}{2\pi^2} \left( \frac{2\pi}{\beta_0} \right)^2 D^E (\lambda), $$ which however does not change appreciably  the integral (\ref{5}), so that one can write
\be  \sigma_f =\frac12 \int d^2 z D^E (z), ~~D^E(z) \cong a e^{-M|z|}, ~~ M\equiv M_0^{(2g)},\label{b1}\ee
where $a$ according to (\ref{9}) is
\be a=16 \pi^2 \alpha^2_s (M) \cdot 0.4 \sigma^2_f.\label{b2}\ee

From (\ref{b1}) and (\ref{b2}) one obtains the  self-consistency  condition
\be  \sigma_f \geq 16 \pi^2 \cdot 0.4 \sigma_f^2 \alpha^2_s (M) \frac{\pi}{M^2}, \label{b3}\ee
where one should take into account the relation
\be \alpha_s (M) = \frac{4\pi}{\beta_0 ln \left( \frac{M^2+M^2_B}{\Lambda^2}
\right)}\label{b4}\ee
with $M_B \cong \sqrt{2\pi\sigma_f}\approx 1$ GeV, $M=2$ GeV,   where we take into account the IR regularization of $\alpha_s$ \cite{84}.  As a result one obtains from (\ref{b3}) and (\ref{b4}) the connection of $\Lambda_\sigma$ and $\sigma$
\be \Lambda^2_\sigma = ( M^2 + M^2_B) \exp \left( -\frac{\sqrt{\sigma_f}}{M} 17.7 \frac{\sqrt{1-1/N^2_c}}{\left( 1- \frac{2}{11} \frac{n_f}{N_c}\right)} \right), \label{b5}\ee
and for $\sigma_f =0.18$ GeV, $M=2$ GeV, $M_B =1$ GeV, $N_c\to \infty$ one obtains  $\Lambda_\sigma =0.342$ GeV. Here one must take into account 
 the  difference between the space  constant $\Lambda_\sigma  $ and the momentum  space $\Lambda_{QCD}$, see \cite{90,91} for the  analysis, $\Lambda_\sigma\approx 1.3 \Lambda_{QCD}$, yielding $\Lambda_{QCD} = 0.29$ GeV.

One can check that for $\sigma = 0.18$ GeV$^2$, $M=2$ GeV  one obtains $\alpha_s (2$ GeV)$\approx 0.33$, which is within the  PDG limits,  and both relations (\ref{b3}) and (\ref{b4}) are satisfied. In this way the self-consistency check  shows the reliability of the obtained gluelump string tension.

In addition one check the behavior of $D^E(z)$ in (\ref{12}) $vs$ lattice data for $D^E(z)$ and $D_1^E(z)$, obtained in  \cite{38,39,40,41,42}.

The general structure of these correlators on the lattice can be approximated as $D^E\sim \frac{a}{z^4} + b e^{-z/\lambda}$,  with $\lambda \sim 0.2$ fm  in  resonable agreement with (\ref{12}). As it is discussed in section 4, this lattice behavior  contradicts the field correlators obtained in  DGL, $D^E, D_1^E \sim O\left(  \frac{e^{-z/\lambda}}{z^4}\right)$, see \cite{10,61*}  for more detail.

\end{document}